\documentclass[preprint,12pt]{elsarticle}
\usepackage{multirow}
\usepackage{url}
\usepackage{paralist}
\usepackage[table]{xcolor}
\usepackage[normalem]{ulem}
\usepackage{graphicx, subfigure}
\usepackage{makecell}
\usepackage{amssymb}
\usepackage{acronym}

\usepackage{color}

\newcommand{\sigla}{\textit{Slice Agent}}

\definecolor{mygreen}{rgb}{0.0, 0.5, 0.0} 
\definecolor{Gray}{gray}{0.9}

\journal{Computers \& Electrical Engineering}

\begin{document}

\begin{frontmatter}

\title{Slice Agent: Identifying and Isolating Slices in Shared Open Radio Unit}

\author[unisinos]{Felipe Arnhold}
\ead{felipe.arnhold@outlook.com}

\author[ufg]{Flávio Rocha}
\ead{flaviogcr@ufg.br}

\author[unisinos]{Lucio Prade}
\ead{luciorp@unisinos.br} 

\author[unisinos]{Cristiano Bonato Both}\corref{correspondente}
\ead{cbboth@unisinos.br}

\cortext[correspondente]{Corresponding author}

\address[unisinos]{University of Vale do Rio dos Sinos (UNISINOS), São Leopoldo -- Brazil}

\address[ufg]{Universidade Federal de Goiás (UFG), Goiânia -- Brazil}

\begin{abstract}

\ac{NSaaS} is a key enabler of Beyond \ac{5G} and \ac{6G} networks, supporting next-generation applications such as extended reality (XR), immersive services, and the tactile Internet. These networks must provide native support for slice-aware services across the entire \ac{RAN}, including the \ac{RU}, \ac{DU}, \ac{CU}, and transport segments (fronthaul, midhaul, and backhaul). However, uplink slicing identification in shared \acp{O-RU} presents a fundamental challenge because the \ac{O-DU} handles scheduling, and the \ac{O-RU} does not inherently know which uplink data belongs to which slice. In \ac{MP2MP} fronthaul scenarios, this limitation is further exacerbated by synchronization and timing constraints, which necessitate that the \ac{O-RU} process control messages and the encapsulated data be delivered with ultra-low latency. To address this issue, we propose a slicing agent embedded in the \ac{O-RU} that identifies slices and segregates uplink data into slice-specific \ac{eCPRI} packets. Our design employs a pipeline architecture with dedicated paths for time-sensitive, flexible slicing, enabling slice isolation and prioritization. When implemented on an \ac{FPGA}, the agent processes each packet in 2 clock cycles, supporting up to 3822 slices per slot. Experimental results validate the approach, showing its feasibility, scalability, and high-performance capabilities for real-time, slice-aware uplink processing in Beyond \ac{5G} and \ac{6G} Open RAN deployments.

\end{abstract}


\end{frontmatter}

\acrodef{5G}{Fifth Generation}
\acrodef{6G}{Sixth Generation}
\acrodef{3GPP}{Third Generation Partnership Project}
\acrodef{AMF}{Access and Mobility Management Function}
\acrodef{BBU}{Baseband Unit}
\acrodef{CI/CD}{Continuous Integration/Continuous Delivery}
\acrodef{CSMF}{Communication Service Management Function}
\acrodef{D-RAN}{Distributed RAN} 
\acrodef{eMBB}{Enhanced Mobile Broadband}
\acrodef{IETF}{Internet Engineering Task Force}
\acrodef{IEEE}{Institute of Electrical and Electronics Engineers}
\acrodef{IaC}{Infrastructure as Code}
\acrodef{IoT}{Internet of Things}
\acrodef{LTE}{Long Term Evolution}
\acrodef{mMTC}{Massive Machine-Type Communication}
\acrodef{NFV}{Network Functions Virtualization}
\acrodef{NASP}{Network Slice as a Service Platform} 
\acrodef{NEST}{Network Slice Type}
\acrodef{NG-RAN}{Next Generation Radio Access Network}
\acrodef{NS}{Network Slicing}
\acrodef{NSaaS}{Network Slice as a Service}
\acrodef{NSC}{Network Slice Controller}
\acrodef{NSMF}{Network Slice Management Function}
\acrodef{NSSMF}{Network Slice Subnet Management Function}
\acrodef{NR}{New Radio}
\acrodef{NRF}{Network Repository Function}
\acrodef{SBA}{Service-Based Architecture}
\acrodef{SLA}{Service Level Agreement}
\acrodef{SLS}{Service Level Specification}
\acrodef{SDN}{Software-Defined Networking}
\acrodef{SMF}{Session Management Function}
\acrodef{UPF}{User Plane Function}
\acrodef{UDM}{Unified Data Management}
\acrodef{UE}{User Equipment}
\acrodef{URLLC}{Ultra-Reliable Low Latency Communications}
\acrodef{VPN}{Virtual Private Network}
\acrodef{GSMA}{Global System for Mobile Communications Association}
\acrodef{ZSM}{Zero-Touch Network and Service Management}
\acrodef{5GC}{5G Core}
\acrodef{B5G}{Beyond Fifth Generation}
\acrodef{ACM}{Association for Computing Machinery}
\acrodef{ONAP}{Open Network Automation Platform}
\acrodef{OSM}{Open Source MANO}
\acrodef{VNF}{Virtual Network Function}
\acrodef{OpenSlice}{OpenSlice Framework}
\acrodef{PLMN-ID}{Public Land Mobile Network IDentifier}
\acrodef{SST}{Slice Service Type}
\acrodef{SD}{Slice Differentiator}
\acrodef{AI}{Artificial Intelligence}
\acrodef{E2E}{End-to-End}
\acrodef{DLT}{Distributed Ledger Technologies}
\acrodef{NSI}{Network Slice Instance}
\acrodef{ETSI}{European Telecommunications Standards Institute}
\acrodef{O-RAN}{Open \ac{RAN}}
\acrodef{RN1}{Radio Access Network Interface}
\acrodef{TN1}{Transport Network Interface}
\acrodef{CN1}{Core Network Interface}
\acrodef{NEF}{Network Exposure Function}
\acrodef{SMO}{Service Management and Orchestration}
\acrodef{Free5gc}{Free 5G Core}
\acrodef{ONOS}{Open Network Operating System}
\acrodef{JSON}{JavaScript Object Notation}
\acrodef{NSST}{Network Slice Templates}
\acrodef{NST}{Network Slice Template}
\acrodef{YAML}{YAML Ain't Markup Language}
\acrodef{HTTP}{Hypertext Transfer Protocol}
\acrodef{VLAN}{Virtual Local Area Network}
\acrodef{REST}{Representational State Transfer}
\acrodef{API}{Application Programming Interface}
\acrodef{IP}{Internet Protocol}
\acrodef{KubeAPI}{Kube-API Server}
\acrodef{Helm}{Helm Chart Package Manager}
\acrodef{Mininet}{Mininet Virtual Network Emulator}
\acrodef{5G}{Fifth Generation}
\acrodef{Calico}{Calico}
\acrodef{Multus}{Multus}
\acrodef{non3GPP}{Non-3GPP}
\acrodef{IPSec}{Internet Protocol Security}
\acrodef{OCloud}{O-Cloud Environment}
\acrodef{RAN}{Radio Access Network}
\acrodef{NF}{Network Function}
\acrodef{TN}{Transport Network}
\acrodef{vCPU}{Virtual CPU}
\acrodef{RAM}{Random Access Memory}
\acrodef{CPU}{Computer Processing Unit}
\acrodef{S-NSSAI}{Single Network Slice Selection Assistance Information}
\acrodef{ITU-T}{International Telecommunication Union Telecommunication Standardization Sector}
\acrodef{K8s}{Kubernetes}
\acrodef{CGROUPS}{Control Groups}
\acrodef{PDU}{Protocol Data Unit}
\acrodef{GTP-5G}{GPRS Tunneling Protocol for 5G}
\acrodef{5G-TOURS}{5G-TOURS Project}
\acrodef{5GZORRO}{5GZORRO Project}
\acrodef{5Growth}{5Growth Project}
\acrodef{5G-COMPLETE}{5G-COMPLETE Project}
\acrodef{PCF}{Policy Control Function}
\acrodef{AUSF}{Authentication Server Function}
\acrodef{NSSF}{Network Slice Selection Function}
\acrodef{NWDAF}{Network Data Analytics Function}
\acrodef{NEF}{Network Exposure Function}
\acrodef{QoS}{Quality of Service}
\acrodef{NG-RAN}{Next Generation \ac{RAN}}
\acrodef{RU}{Radio Unit}
\acrodef{DU}{Distributed Unit}
\acrodef{CU}{Central Unit}
\acrodef{CN}{Core Network}
\acrodef{NFVI}{Network Functions Virtualization Infrastructure}
\acrodef{MANO}{Management and Orchestration}
\acrodef{DRL}{Deep Reinforcement Learning}
\acrodef{DCAE}{Data Collection, Analytics, and Events}
\acrodef{CNN}{Convolution Neural Network}
\acrodef{RNN}{Recurrent Neural Network}
\acrodef{DQN}{Deep Q-Network}
\acrodef{CNF}{Cloud-native Network Function}
\acrodef{KPI}{Key Performance Indicator}
\acrodef{RIC}{RAN Intelligent Controller}
\acrodef{RL}{Reinforcement Learning}
\acrodef{RRH}{Remote Radio Head}
\acrodef{GAN}[GAN]{Generative Adversarial Network}
\acrodef{LSTM}[LSTM]{Long Short-term Memory}
\acrodef{SFaaS}{Slice Federation-as-a-Service}
\acrodef{SFC}{Service Function Chain}
\acrodef{NFVO}{Network Function Virtualization Orchestrator}
\acrodef{NSD}{Network Slice Descriptor}
\acrodef{GST}{Generic Network Slice Template}     
\acrodef{TMF}{TeleManagement Forum}
\acrodef{gNB}[gNB]{next-generation Node B}
\acrodef{WAN}[WAN]{Wide Area Network}
\acrodef{UTC}[UTC]{Coordinated Universal Time}
\acrodef{VM}[VM]{Virtual Machine}
\acrodef{mMTC}[mMTC]{massive Machine-Type Communications}
\acrodef{NSSAI}[NSSAI]{Network Slice Selection Assistance Information}
\acrodef{SCTP}[SCTP]{Stream Control Transmission Protocol}
\acrodef{N3IWF}[N3IWF]{Non-3GPP Interworking Function}
\acrodef{LLS}[LLS]{Lower Layer Split}
\acrodef{O-RU}[O-RU]{Open-RU}
\acrodef{O-DU}[O-DU]{Open-DU}
\acrodef{O-CU}[O-CU]{Open-CU}
\acrodef{SCF}[SCF]{Small Cell Forum}
\acrodef{HLS}[HLS]{High Layer Split}
\acrodef{MAC}[MAC]{Medium Access Control}
\acrodef{CO-DBA}[CO-DBA]{Cooperative Dynamic Bandwidth Allocation}
\acrodef{DBA}[DBA]{Dynamic Bandwidth Allocation}
\acrodef{PON}[PON]{Passive Optical Network}
\acrodef{OLT}[OLT]{Optical Line Terminal}
\acrodef{ASIC}[ASIC]{Application-Specific Integrated Circuit}
\acrodef{FPGA}[FPGA]{Field-Programmable Gate Array}
\acrodef{COTS}[COTS]{Commercial Off-The-Shelf}
\acrodef{MP2MP}[MP2MP]{MultiPoint-to-MultiPoint}
\acrodef{PHY}[PHY]{Physical Layer}
\acrodef{eCPRI}[eCPRI]{enhanced Common Public Radio Interface}
\acrodef{PRB}[PRB]{Physical Resource Block}
\acrodef{I/Q}[I/Q]{In-phase/Quadrature}
\acrodef{FIFO}[FIFO]{First In, First Out}
\acrodef{MTU}[MTU]{Maximum Transmission Unit}
\acrodef{AMD}[AMD]{Advanced Micro Devices}
\acrodef{UART}[UART]{Universal Asynchronous Receiver Transmitter}
\acrodef{SUT}[SUT]{System Under Test}
\acrodef{LUT}[LUT]{Look-up Table}
\acrodef{FF}[FF]{Flip-Flop}
\acrodef{BRAM}[BRAM]{Block RAM}
\acrodef{DSP}[DSP]{Digital Signal Processing}
\acrodef{ITU}[ITU]{International Telecommunication Union}
\acrodef{TDM}[TDM]{Time-Division Multiplexing }
\acrodef{WG1}[WG1]{Work Group 1}
\acrodef{WG4}[WG4]{Work Group 4}
\acrodef{CUS}[CUS]{Control, User, and Synchronization Plane}
\section{Introduction}


\acf{5G} mobile networks have adopted new technologies, such as network slicing and disaggregated \acf{RAN}, to deliver services with different requirements, including \ac{eMBB}, \ac{URLLC}, and \ac{mMTC} \cite{gsma.5gGuide.2019}. At first, mobile operators used network slices in a \emph{static} way because there were only a few slices, and orchestration methods were new. As demand for specialized services and \ac{ZSM} increased, research and open-source efforts began to enable \emph{dynamic} network slicing. \ac{ETSI} forecasts that soon, \acf{NSaaS} will be provided across several domains, including \ac{RAN}, transport, and core networks \cite{etsi.zsm.2021}. In Beyond \ac{5G} and \acf{6G} networks, \ac{NSaaS} is expected to be a key, built-in part of the system, enabling \ac{E2E} service access for new use cases such as immersive experiences, extended reality, and the tactile Internet \cite{etsi.zsm.2021}.

Disaggregated \acp{RAN} were conceptualized more than ten years ago, dividing the monolithic \ac{D-RAN} into \acp{RRH} and \acp{BBU} to reduce costs and enhance \ac{QoS} for mobile operators \cite{ZANFERRARIMORAIS.2020}. A centralized \ac{BBU} pool gave rise to Cloud-\ac{RAN}, leveraging cloud computing principles. The \ac{3GPP} introduced a novel segmentation in Release 15 \cite{3gpp.38.300.2018}, further refined in Releases 16 \cite{3gpp.38.300.r16.2023} and 17 \cite{3gpp.38.300.r17.2023}, by dividing the \ac{RAN} into three elements: \acf{RU}, \acf{DU}, and \acf{CU}. While such functional splits and partial disaggregation are already present in current \ac{5G} deployments, their large-scale openness, full programmability, and tight integration with slicing mechanisms are still evolving. The \ac{O-RAN} ALLIANCE was established to create open interfaces for the cohesive operation of these elements, independent of vendors, and to facilitate the development of new services and applications \cite{oran.oad.wg1.2023}. In this context, Beyond \ac{5G} and \ac{6G} networks are expected to extend and consolidate these disaggregation principles, incorporating native support for \ac{NSaaS} and programmable \ac{RAN} architectures, tightly integrating slice services with the \ac{RU}, \ac{DU}, and \ac{CU}, as well as with fronthaul, midhaul, and backhaul transport connections.

In the \ac{O-RAN} architecture, the chosen \ac{LLS} aims to propose a good trade-off between the required bandwidth and the complexity of the \ac{O-RU}, reducing costs \cite{oran.cusFront.wg4.2023}. From the \ac{O-RAN} specifications, \acf{O-DU} and \ac{O-CU} already support slicing, but not \acf{O-RU} \cite{oran.slice.wg1.2023}. Slicing support adds functionality to the \ac{O-RU}, and maintaining a relatively low cost remains challenging. These \acp{LLS} become more challenging when the shared \ac{O-RU} and multipoint-to-multipoint fronthaul scenario is considered, as new challenges involving synchronization and slice identification must be addressed \cite{oran.usecases.wg1.2023}.
The main challenge in identifying slices within the \ac{O-RU} lies in the uplink process, because the \ac{O-RU} does not know which data should be sent to the \ac{O-DU}, as it is the \ac{O-DU} that performs the scheduling process \cite{oran.oad.wg1.2023}. In this context, the \ac{O-RU} must receive precise uplink scheduling information in advance, enabling it to associate the received signals from \acp{UE} with the corresponding slices, encapsulate them, and forward them to the appropriate \ac{O-DU}. This process is subject to strict time constraints, requiring the \ac{O-RU} to be synchronized with other network components and to process scheduling information within a short time window \cite{oran.cusFront.wg4.2023}.

We propose an architecture for slicing identification and segregation in the uplink process, addressing this challenge directly within the \ac{O-RU}, denoted \sigla{}. Our proposal uses the existing \ac{O-RAN} fronthaul to receive slice information and encapsulate radio data into separate Ethernet packets for each slice. Additionally, it designs two processing units to enable slice isolation and prioritization. Moreover, we present an experimental case to evaluate the performance of the \sigla{}. The main contributions of this work are:

\begin{itemize}
    \item[($i)$] We present \sigla{}, an architecture designed to enable slicing support in the uplink process within the \ac{O-RU}.
    \item[($ii)$] We introduce a pipeline and parallel architecture to achieve high performance in slicing identification.
    \item[($iii)$] We demonstrate high processing capacity, capable of handling the maximum theoretical load in specific scenarios.
    \item[($iv)$] We develop a simulation environment to validate \sigla{} and carry out scalability tests using the real prototype.
\end{itemize}

The results demonstrate the feasibility and high performance of the proposed architecture for real-time, slice-aware uplink processing. By leveraging a pipeline design with dedicated paths for time-sensitive, flexible \ac{RAN} slicing, the \acf{FPGA}-based implementation can process each packet in 2 clock cycles, achieving ultra-low latency. The system supports up to 3,822 slices per slot across two slicing types, including \ac{URLLC} and \ac{eMBB}, demonstrating substantial scalability under demanding traffic conditions. Experimental validation confirms the approach's effectiveness, highlighting its potential to enable efficient, programmable slicing mechanisms in Beyond \ac{5G} and \ac{6G} networks.

The article is structured as follows: Section~\ref{sec:background} covers the study's essential concepts. Section~\ref{sec:relatedworks} discusses previous research on slicing support in \acp{RU} and fronthaul. Section~\ref{sec:prop} details the \sigla{} architecture. Section \ref{sec:metrics} outlines the prototype implementation and provides an experimental case study to evaluate the performance of the proposed architecture. Lastly, Section~\ref{sec:conc} concludes the article and outlines future directions.
\section{Background}
\label{sec:background}

This section presents fundamental concepts regarding \acp{RAN} and their relevant characteristics as applicable to this article. First, the basic architecture of a mobile network is given, followed by a brief explanation of \ac{RAN}, with a focus on its disaggregation. Finally, the concept of network slicing is introduced. 


\subsection*{\textbf{Mobile network}}


Mobile cellular networks constitute the foundation of contemporary wireless communications, supporting a wide range of services. A cellular network architecture is traditionally structured into three main components: the Core Network, \ac{RAN}, and \ac{UE}. The Core Network, typically deployed in operators’ data centers, is responsible for essential control and user-plane functions, including authentication, mobility management, session establishment, and external network connectivity. \ac{RAN} provides the wireless interface between the \ac{UE} and the Core, performing radio signal processing, resource management, and protocol stack functions required to ensure reliable and efficient communication. \ac{UE} refers to the end devices accessing the network, encompassing smartphones and \ac{IoT} terminals \cite{dahlman20205g,3gpp_ts23501_r20,3gpp_ts23502_r20}. 

This architecture has continuously evolved across successive generations, driven by increasingly stringent requirements for data rate, latency, reliability, and service diversity. In this context, \ac{5G} has reached a mature deployment stage worldwide, with current efforts primarily focused on performance optimization, network automation, and the support of heterogeneous services. At the same time, research initiatives and early standardization activities have increasingly shifted toward defining 6G, aiming to address emerging requirements and extend the architectural foundations established by \ac{5G} \cite{saad2019vision, lin2025bridge}.

\subsection*{\textbf{Radio Access Network}}


\ac{RAN} is a critical component of \ac{5G} networks, connecting \acp{UE}, such as smartphones, to the Core Network and enabling wireless communication and data exchange. In earlier generations of mobile networks, the \ac{RAN} was typically implemented as a monolithic entity, with radio frequency, baseband processing, and control functionalities tightly integrated into a single vendor-dependent device. With the evolution of cellular systems, particularly during the \ac{LTE} era, the first steps toward architectural disaggregation were introduced through the separation of radio and baseband functions. This disaggregation led to the emergence of Cloud-\ac{RAN} architectures, in which the radio functionality was implemented by \acp{RRH} deployed near the antennas, while the \acp{BBU} were centralized to enable resource pooling and coordination. Subsequent advances in network virtualization further enabled the \ac{BBU} to be a software-based entity running on cloud infrastructures \cite{checko2014cloud, svLarsen2018}. Building upon these developments, the stringent and heterogeneous requirements introduced by \ac{5G}, including low latency, high data rates, and massive connectivity, have driven the adoption of more flexible and scalable \ac{RAN} architectures. In this context, the \ac{RAN} has evolved toward a fully disaggregated model, formally structured into \ac{CU}, \ac{DU}, and \ac{RU}, where the \ac{CU} and \ac{DU} conceptually extend the traditional \ac{BBU}, and the \ac{RU} represents the evolution of the \ac{RRH} \cite{3gpp_ts38401_r19}.

In the context of \ac{5G}, functional splits are crucial. Functional splits refer to how the \ac{RAN}'s processing tasks are divided among the \ac{CU}, \ac{DU}, and \ac{RU}. Therefore, the choice of split impacts the complexity of each unit and the bandwidth and latency requirements for each network segment. The \ac{3GPP}, which defines global telecommunications standards, introduced eight functional split options, each tailored to different network requirements and deployment scenarios.

\begin{figure}[!h]
    \centering
        \includegraphics[width=0.9\textwidth]{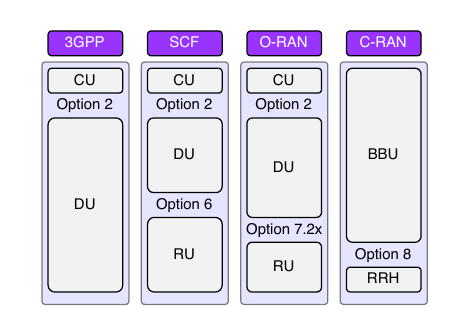}
    \caption{Functional split options by different organizations, compared with a C-RAN.}
    \label{fig:spliResume}
\end{figure}

The \ac{HLS}, which separates the \ac{CU} and \ac{DU}, was defined by \ac{3GPP} as Option 2 and is considered to offer the best cost-benefit ratio. This split balances processing capabilities and network efficiency, making it a widely adopted standard in \ac{5G} deployments. On the other hand, the \ac{LLS}, which separates the \ac{DU} and \ac{RU}, was not explicitly defined by \ac{3GPP}. This decision was made because the optimal choice for the \ac{LLS} depends heavily on specific applications and deployment scenarios. As a result, different industry groups have proposed various options for this split, such as Split 6 by the \ac{SCF} and Split 7.2x by the \ac{O-RAN} Alliance, each tailored to meet the distinct needs of \ac{5G} networks \cite{PlaceRAN.2023}. Figure \ref{fig:spliResume} presents an overview of the division of the New Radio protocol into \ac{CU}, \ac{DU}, and \ac{RU} in several architectures.

The disaggregation of the \ac{RAN} into \ac{CU}, \ac{DU}, and \ac{RU} introduces new interfaces within the network. The backhaul is the interface between the \ac{CU} and the Core network, ensuring connectivity and communication between the \ac{RAN} and the network's central components. The midhaul is the interface between the \ac{CU} and \ac{DU}, while the fronthaul is the interface between the \ac{DU} and \ac{RU}. These interfaces are fundamental to maintaining the flexibility and scalability of 5G networks, enabling operators to optimize the placement and operation of each RAN component for specific use cases and deployment scenarios \cite{arnhold.2024}.

\subsection*{\textbf{Network slicing}}




Network slicing constitutes one of the architectural pillars of \ac{5G}, enabling the deployment of multiple logically isolated networks over a shared physical infrastructure. Instead of statically partitioning resources, slicing allows operators to dynamically tailor connectivity profiles according to heterogeneous service demands, including throughput, latency, reliability, and isolation requirements \cite{ojaghi2019sliced,ojaghi2022impact}. This concept introduces significant orchestration complexity, since diverse service instances must coexist while competing for common computational, transport, and radio resources. However, once this complexity is addressed, network slicing enables operators to allocate network resources more efficiently and deliver customized services to different user types and applications.

From a standardization perspective, the slicing framework defined by \ac{3GPP} establishes the concept of a \ac{NSI}, whose lifecycle is managed through dedicated management functions such as the \ac{NSMF} \cite{3gpp.28.530}. Building upon this foundation, the O-RAN Alliance integrates slicing support into its service management and orchestration layer through the \ac{SMO}, extending slice management capabilities to an \ac{E2E} approach \cite{oran.slice.wg1.2023,3gpp20173gpp}. Complementarily, the IETF specifies a non-prescriptive reference architecture for network slicing in which customer requests for Network Slice Services are handled by a \ac{NSC}, establishing a common framework that promotes interoperability and conceptual alignment across diverse slicing implementations \cite{farrel2024framework}.

The \ac{E2E} slicing nature is a fundamental requirement for this paradigm, since a slice must not be confined to a single domain but span the \ac{5GC}, the \ac{TN}, the \ac{RAN}, and ultimately the \ac{UE}. Such \ac{E2E} continuity ensures that performance guarantees are consistently enforced throughout the entire communication path, thereby enabling \ac{SLA} compliance across heterogeneous network segments \cite{3gpp_ts23501_r20, 3gpp_ts23502_r20}. Consequently, all involved domains must support slice-aware control, monitoring, and resource isolation mechanisms.

Slice identification and traffic differentiation rely on standardized identifiers. For each operator, uniquely defined by the \ac{PLMN-ID}, slice selection during \ac{PDU} session establishment is performed using the \ac{S-NSSAI}, which comprises the \ac{SST} and the \ac{SD} \cite{oran.slice.wg1.2023}. This mechanism allows multiple slices to coexist within the same public land mobile network while maintaining traffic separation and \ac{QoS} assurance. In the \ac{TN}, \ac{PDU} sessions are transported over Ethernet-based infrastructures \cite{oran.mFront.wg4.2023}, and operators must correlate session flows with their corresponding \acp{NSI} to preserve isolation and performance differentiation.
Moreover, slicing supports multi-operator deployments through \ac{RAN} sharing scenarios. Thanks to functional disaggregation and open interfaces, infrastructure components, potentially from different vendors, can be shared among multiple operators while maintaining slice-level separation \cite{oran.slice.wg1.2023}. This capability significantly enhances the flexibility of infrastructure utilization and deployment.

Beyond static slice provisioning, the slicing paradigm is evolving toward a service-oriented model commonly referred to as \ac{NSaaS}. In this approach, slices are instantiated, scaled, and terminated on demand, similarly to cloud-native services. Such a model enables vertical industries and enterprise customers to request customized connectivity solutions aligned with specific business objectives, requiring advanced automation, programmability, and lifecycle management across orchestration frameworks \cite{foukas2017orion}. The realization of \ac{NSaaS} therefore depends on tight integration between slice management functions, virtualization platforms, and standardization for all architectural domains.

\section{Related work}
\label{sec:relatedworks}

Several studies have explored tools to support slicing in \ac{5G} networks. However, there is a noticeable gap in the studied technology, particularly regarding slicing support directly by the \ac{RU} \cite{arnhold.2024, Ornonez.2021}. The survey presented by Arnhold \textit{et al.} \cite{arnhold.2024} provides a detailed review of the work in this field, which identified opportunities that have led to the development of the technology presented in this article. In this context, some existing work addresses this concept, primarily focusing on fronthaul. 

The 5G Crosshaul project \cite{_A16, _A18, _A19} defines an architecture for a 5G network that supports crosshaul, characterized by the coexistence of backhaul, midhaul, and fronthaul interfaces within a single physical network. For this coexistence to be achieved, the system must inherently support network slicing to ensure traffic isolation. In the fronthaul case, where slicing identification is unavailable, slicing support was provided by dedicated network elements deployed within the transport network between devices, enabling proper traffic routing and prioritization. Under this approach, new and customized transport elements will be required, increasing the cost of the network infrastructure. A slicing technology provided directly by an \ac{RU} eliminates the need for such additional elements, allowing the use of existing infrastructure. This approach enables the \ac{RU} to manage multiple slices independently, simplifying the network architecture and reducing the complexity associated with routing and split selection. By integrating slicing capabilities directly into the \ac{RU}, it is possible to support diverse slices within a shared \ac{RU} scenario, leveraging existing infrastructure without requiring significant modifications or additional network components.

The Sliced \ac{RAN} project \cite{_A5, _A5.1} also presents an interesting proposal to support slicing \ac{E2E}. In this case, the fronthaul is designed to support multiple slices, ensuring each slice receives sufficient bandwidth. This objective is achieved by determining the optimal routing path across the available network elements and selecting the optimal functional split for each slice. However, this project assumes that an \ac{RU} can serve only one \ac{DU}, limiting its applicability in scenarios involving a shared \ac{RU}. In such cases, where a single \ac{RU} is expected to support multiple slices originating from different \acp{DU}, this project falls short, making it unsuitable for the shared \ac{RU} use case.

In the New Radio protocol, for a downlink or uplink transmission to occur, it must first be scheduled by the scheduling algorithm running in the \ac{MAC} layer within the \ac{DU}. This scheduling process provides advanced knowledge of downlink and uplink information, enabling informed decisions and reducing latency. The \ac{CO-DBA} is an \ac{ITU} standard-supported technique (G.Suppl.71) that reduces upstream latency in \ac{TDM}-\acp{PON}, particularly for \ac{5G} mobile fronthaul \cite{ITU-G-Sup71-2023}. In this case, \ac{CO-DBA} calculates the necessary bandwidth required by \acp{PON} to perform downlink and uplink transmissions for a specific slice \cite{_A13}. \ac{CO-DBA} can determine the required bandwidth by obtaining scheduling information in advance, even before the actual data transfer begins, reducing latency and supporting low-latency applications, a key goal of beyond \ac{5G} and \ac{6G} networks. Several works utilize the \ac{CO-DBA} technology, as presented by Uzawa \textit{et al.} \cite{_A13}, which proposed a \ac{CO-DBA} algorithm for a time-division multiplex \ac{PON}. In this work, the time interval is divided into multiple grant cycles to ensure guaranteed bandwidth allocation. These grant cycle divisions are done using the scheduling information received in advance. The same concept is used by Das \textit{et al.} \cite{_A2, _A7, _A10, _A11} and Mondal and Rufini \cite{_A4}. In all cases, the \ac{CO-DBA} is used to ensure bandwidth allocation to slicing with different requirements in a low-latency scenario. However, the shared \ac{RU} use case, in which a single \ac{RU} supports multiple slices from different \acp{DU}, is not addressed by this approach, leaving a gap in current technology.

Budhdev \textit{et al.} \cite{_A8} used a similar technique to identify a slice. The authors proposed a method implemented in a programmable switch that receives scheduling information in advance, enabling accurate identification and forwarding of incoming data packets. The focus was on the uplink procedure, as the \ac{RU} lacked information about the incoming data. However, programmable switches are costly, and in a complex \ac{RAN} with interface coexistence and multiple \acp{RU} and \acp{DU}, synchronization between all transport elements poses a significant challenge. In fact, synchronization across fronthaul transport elements is a challenge in disaggregated and multipoint \ac{RAN} architectures, affecting not only the solution proposed by Budhdev \textit{et al.}, but also any approaches that require tight timing coordination among \acp{RU}, \acp{DU}, and transport components. 



In the current specifications of the O-RAN Alliance regarding the fronthaul, the focus is primarily on functional disaggregation and the definition and consolidation of an open interface, namely Open Fronthaul, between the \ac{O-DU} and the \ac{O-RU}, as described in the latest release of the \ac{WG4} \ac{CUS} specifications~\cite{oran_wg4_cus_v20}. \ac{WG4} is the main initiative within the O-RAN architecture for Open Fronthaul. The document released by this group spans over 500 pages and provides detailed specifications for architecture, transport, security, and the control, user, and synchronization planes. However, despite this level of detail, support for network slicing is not explicitly addressed at the fronthaul level.

The O-RAN Alliance's vision of network slicing has evolved incrementally, reflecting the broader goal of extending slicing capabilities beyond the core network and logical \ac{RAN} functions into the transport domain. This evolution is more clearly articulated in \ac{WG1}, which defines the slicing architecture, use cases, and requirements for O-RAN systems~\cite{oran_wg1_slicing_v14_2025}. In particular, slicing is treated as an \ac{E2E} concept, encompassing core, RAN, and transport segments.
However, a clear mismatch emerges between these two perspectives. While \ac{WG1} defines slicing as an \ac{E2E} characteristic, the fronthaul specifications in \ac{WG4} lack explicit mechanisms to support slice-aware transport. As a result, slicing in the fronthaul is implicitly assumed to rely on traditional \ac{QoS} differentiation and traffic prioritization mechanisms, rather than on native slice identification and isolation.

This limitation becomes even more critical in emerging deployment scenarios, such as neutral-host and multi-operator \acp{RAN}, where a common physical infrastructure must support multiple tenants with distinct service requirements. Within this context, the use cases defined by \ac{WG4}, especially Use Case~7 (multi-slice support) and Use Case~10 (multi-operator and multi-vendor scenarios), foresee that a single \ac{O-RU} will simultaneously serve multiple slices and multiple \acp{O-DU}~\cite{oran_wg1_slicing_v14_2025, oran_wg4_cus_v20}. This capability inherently requires slice-aware traffic handling in the fronthaul transport plane.
Therefore, a key challenge arises from the absence of native slice identifiers in current Open Fronthaul transport protocols, such as \ac{eCPRI}. While the 3GPP defines slice identifiers, such as \ac{S-NSSAI}, at the system level~\cite{3gpp_ts23501_r20,3gpp_ts23502_r20}, these identifiers are not directly available at the fronthaul transport layer. This specification creates a gap between the slice definition and the slice deployment, particularly in the transport segment closest to the O-RU.

To overcome this limitation, current discussions point toward the need for mapping higher-layer slice-related information, such as \ac{PLMN-ID} or \ac{QoS} profiles, into transport-level identifiers. One promising approach leverages Ethernet-based mechanisms, such as VLAN tagging (IEEE 802.1Q), priority code points (PCP), and, in some cases, IP/UDP flow descriptors, to differentiate traffic from distinct slices. By mapping \ac{PLMN-ID} or slice-related attributes to VLAN IDs and associated priority levels, it becomes possible to enforce traffic separation across both user and control planes in the fronthaul.
Therefore, the lack of explicit slice-aware mechanisms in the fronthaul represents a fundamental limitation for achieving true \ac{E2E} network slicing. In particular, it can become a bottleneck for strict \ac{QoS} enforcement, especially in latency-sensitive applications such as industrial automation and mission-critical communications~\cite{rocha2026optimal}. This gap underscores the need for mechanisms that bring slice awareness into the fronthaul transport domain.

In the following, we analyze the related work from a comparative perspective across three architectural dimensions: crosshaul capability, transport specialization requirements, and support for \ac{MP2MP} connectivity. This structured comparison highlights common design assumptions and clarifies the architectural positioning of the proposed \sigla{}.
Table \ref{tab:crosshaul_features} presents a comparison of the related work. Most proposals support a crosshaul network. For example, the 5G Crosshaul project \cite{_A16, _A18, _A19} achieved the state-of-the-art with a packet-oriented network. These solutions used techniques such as \ac{VLAN} and \ac{IP} to identify and isolate each interface. Additionally, other works utilized optical networks, which can support multiple domains by isolating them in the time and frequency domains. Since the proposed \sigla{} uses the \ac{O-RAN} fronthaul architecture, which is also a packet-oriented network, it is possible to support multiple interfaces on the same network, isolating each with \acp{VLAN}, as in other projects.
However, the proposals by Budhdev \textit{et al.} \cite{_A8} and the Sliced RAN project \cite{_A5, _A5.1} considered only fronthaul data, limiting their functionality to a single transport interface and not supporting a complete crosshaul network.

Most existing studies focus on the transport domain and rely on specialized transport elements, which increases overall network complexity. For instance, in the 5G Crosshaul project, specialized routers handle fronthaul routing. In \cite{_A13}, a dedicated \ac{DBA} must be implemented within an optical element, and in the EAST-WEST \ac{PON} architecture, a specialized optical router enables dynamic horizontal-plane routing rather than the traditional north-south routing. Similarly, Mondal and Ruffini \cite{_A4} developed a custom server to facilitate data exchange between two \ac{PON} stages. The use of programmable switches by Budhdev \cite{_A8} increased the implementation’s cost and complexity, requiring a distinct algorithm on each network switch. However, the proposed \sigla{} system enables slicing identification and isolation at the \ac{RU}, which assigns \acp{VLAN} to designate routes, allowing the use of commercial switches with \ac{VLAN} capabilities to interface with \ac{RAN} elements. Ojaghi \textit{et al.} \cite{_A5, _A5.1} used the same approach, in which the fronthaul model relies on simple transport elements.
In the shared \ac{RU} context, only the proposed \sigla{} is capable of handling this task. By using scheduling information in advance, the \textit{Slice Agent} is designed to separate the \ac{RU} data into several packets, each of which can be directed to a different \ac{DU}. By identifying these packets with a \ac{VLAN}, the transport switching elements can forward the data along the correct path. Although the approaches proposed by Ojaghi \textit{et al.} \cite{_A5, _A5.1} support dynamic reconfiguration of links between elements, they assume a single \ac{RU}-to-\ac{DU} connection, thus preventing one \ac{RU} from simultaneously serving multiple \acp{DU}.


As a third comparison dimension, we examine support for \ac{MP2MP} architectures. Related work based on packet-switched architectures \cite{_A16, _A18, _A19, _A5, _A5.1, _A8} inherently enables \acf{MP2MP} connectivity.
The specialized WEST-EAST \ac{PON} architecture \cite{_A2, _A7, _A10, _A11} also supports this characteristic. In this architecture, a dedicated optical router is defined, making it possible to reroute the optical signal to a same-level element. Even if the signal is directed to a specific point, the mesh architecture enables dynamic routing, allowing changes not only to the paths but also to the elements connected to them. The \ac{DBA} scheme in \cite{_A13} is an algorithm specifically designed for the \ac{OLT} element in a \ac{PON}, supporting a one-to-n connection rather than a \ac{MP2MP} setup. The proposed \sigla{} leverages packet switching to interconnect multiple \ac{RAN} elements via the fronthaul.

\begin{table}[ht]
    \centering
    \resizebox{\linewidth}{!}{%
    \begin{tabular}{l c c c c}
        \hline \hline
          \textbf{Works} &
          \makecell{\textbf{Crosshaul} \\ \textbf{support}} & 
          \makecell{\textbf{Fronthaul} \\ \textbf{complexity}} & 
          \makecell{\textbf{Shared} \\ \textbf{O-RU}} & 
          \makecell{\textbf{MP2MP} \\ \textbf{architecture}} \\
        \hline \hline
        
        \rowcolor{Gray}5G Crosshaul \cite{_A16, _A18, _A19}                                         & \checkmark    & High  & -             & \checkmark \\ \hline
        
        Sliced RAN \cite{_A5, _A5.1}                                                                & -    & Low   & -             & \checkmark \\ \hline
        
        \rowcolor{Gray} DBA scheme \cite{_A13}                   & \checkmark    & High  & -             & - \\ \hline
        
        \makecell[l]{EAST-WEST PON \\ architecture \cite{_A2, _A7, _A10, _A11}} & \checkmark    & High  & -             & \checkmark \\ \hline
        
        \rowcolor{Gray}Open access-edge server \cite{_A4}                                           & \checkmark    & High  & -             & \checkmark \\ \hline
        
        FSA with programmable switch \cite{_A8}                                                     & -             & High  & -          & \checkmark \\ \hline
        
        \rowcolor{Gray}\sigla{}                                                                  & \checkmark    & Low   & \checkmark    & \checkmark \\
        \hline \hline
    \end{tabular}
    }
    \caption{Related work comparison.}
    \label{tab:crosshaul_features}
\end{table}

The state-of-the-art review reveals that slicing identification and isolation are not performed by the \ac{RU}, and that current techniques focus on the transport domain. This approach increases the cost of transport elements and necessitates specific support for slicing. Transferring slice identification to the \ac{RU} offers significant benefits to the overall architecture. First, the \ac{RU} already includes specialized processors, such as \acp{ASIC} and \acp{FPGA}, so adding slicing identification would have minimal impact on the \ac{RU}'s design. 
Second, by performing slice identification directly at the \ac{RU}, slice-aware information can be explicitly embedded into fronthaul packets, enabling transport switches to apply prioritization, traffic engineering, and \ac{QoS} policies in \ac{MP2MP} scenarios without requiring complex per-flow inspection mechanisms. In this way, transport elements remain slice-aware while leveraging standardized mechanisms for differentiated forwarding. 
Finally, use cases, such as a shared \ac{RU} and multiple paths between \ac{RU} and \ac{DU}, can be dynamically arranged. In this article, we propose the \sigla{} within the \ac{RU}, responsible for identifying the slicing and segregating \ac{RU} data into packets for each slice. The following sections describe the proposed architecture.
\section{\sigla{} architecture} 
\label{sec:prop}

This section introduces the \sigla{} architecture and explains how it achieves slicing identification and \ac{RU} data segregation. We begin this description with key decisions, followed by an overview of the proposed architecture. Finally, we detail each block of the proposed architecture.

\subsection*{\textbf{Key decisions}}

The choice of the \ac{LLS} fundamentally defines the fronthaul performance requirements, as it determines the functional split between \ac{O-RU} and \ac{O-DU}, specifying which 5G \ac{NR} \ac{PHY} processing functions are executed at each unit. This architectural decision directly impacts latency, throughput, synchronization accuracy, and implementation cost. In this context, the \ac{O-RAN} architecture is a trend due to its openness and the selection of the \ac{LLS}. \ac{O-RAN} uses the functional split option 7.2x between \ac{O-RU} and \ac{O-DU}, where only the low \ac{PHY} functions of the \ac{NR} are performed by the \ac{O-RU}. This split enables dynamic bandwidth in the fronthaul with a significantly lower-cost \ac{RU}. Moreover, \ac{O-RAN} specifies the fronthaul, providing tools to enable communication between \ac{O-DU} and \ac{O-RU} in compliance with the specifications, including separation of user, control, synchronization, and management planes. Additionally, the fronthaul specification is packet-switched, increasing its flexibility and optimization opportunities. Considering this context, the proposed \sigla{} was developed within the \ac{O-RAN} architecture.

The biggest challenge in the scenario presented is identifying and segregating fronthaul packets by the \ac{O-RU}. The \ac{O-RU} performs only low \ac{PHY}, meaning all scheduling and data identification functions are performed inside the \ac{O-DU}, i.e., the \ac{MAC} layer. Taking into account the downlink flow, the \ac{O-DU} can efficiently pack New Radio data into tagged fronthaul packets, since it has all the necessary information, including which slice each packet belongs to and which \ac{O-RU} packet it must be sent to the \ac{O-DU}. However, in the uplink flow, the \ac{O-RU} does not implement the necessary mechanisms to identify each slice and to pack the \ac{NR} data with the correct tag. Moreover, \ac{O-RAN} does not specify in its documentation the support for slicing in the fronthaul, although it does provide some tools to enable flow identification. As uplink identification remains an open challenge, this project focuses solely on the uplink process, operating within the \ac{O-RU} process.

\subsection*{\textbf{O-RU \sigla{} architecture}} \label{sec:proposed-arch}

\begin{figure}[!ht]  
    \begin{center}
        \caption{Placement of proposed \sigla{}.}
        \label{fig:simpleProposal}
        \includegraphics[width=1\textwidth]{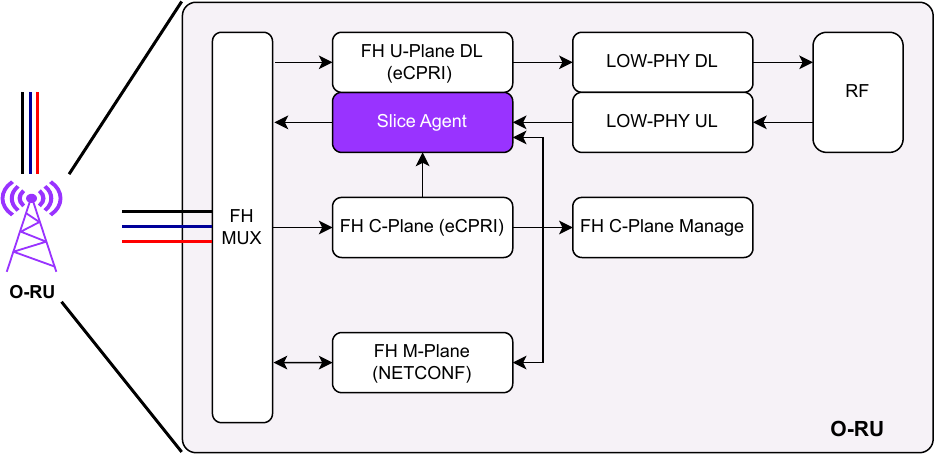}
    \end{center}
\end{figure}

The proposed \sigla{} identifies and encapsulates incoming uplink data in the fronthaul using \acf{eCPRI} over Ethernet. In Figure \ref{fig:simpleProposal}, a simplified architecture of an \ac{O-RU} is presented with the \sigla{} highlighted. The \sigla{} receives uplink data from the LOW-\ac{PHY} processing and communicates with the fronthaul control and management plane to enable the identification of slices and the segregation of RU data.

\begin{figure}[!ht]  
    \begin{center}
        \caption{\ac{O-RU} \sigla{} architecture.}
        \label{fig:proposalArch}
        \includegraphics[width=1\textwidth]{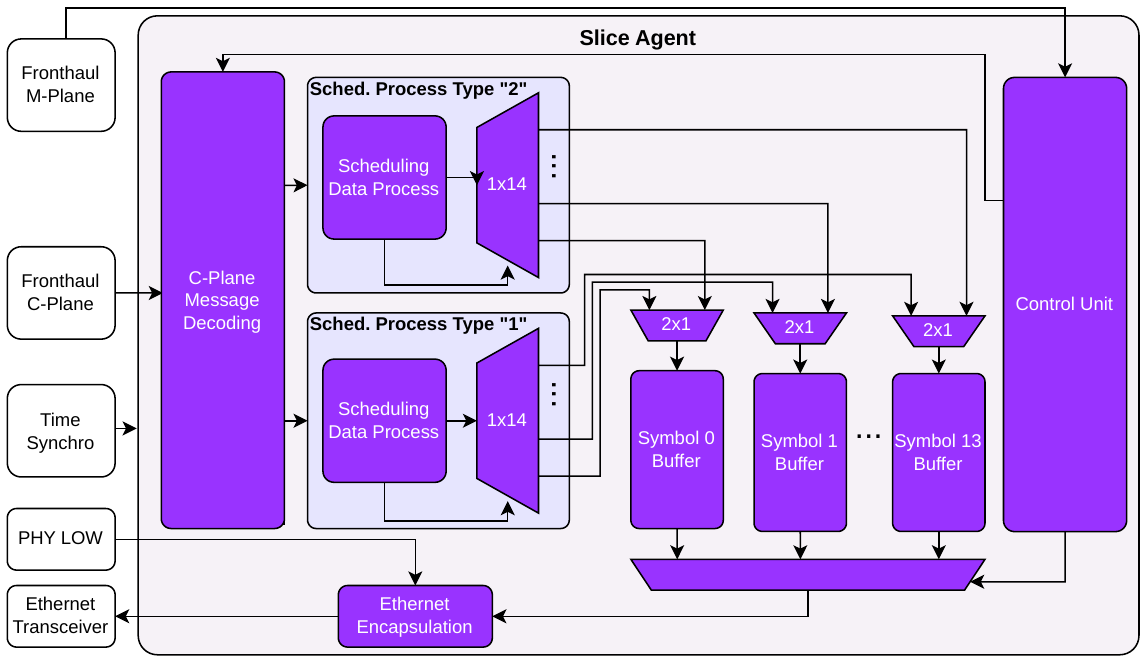}
    \end{center}
\end{figure}

The architecture of the \sigla{} is presented in Figure \ref{fig:proposalArch}. We use the pipeline structure to process scheduling data received via the fronthaul control plane and to encapsulate uplink data from the lower layers of the \ac{O-RU}. In this context, the \sigla{} process can be divided into three main steps. The first step is C-Plane message decoding, which receives and processes the C-Plane message containing each slice's scheduling data. Based on preconfigured parameters, this step analyses the received message and routes it to the appropriate processing unit. 

The second step performs scheduling data processing, transforming slice-oriented information into symbol-oriented information, which is easier to handle by encapsulation. This step comprises two units: one for slices that must be processed as soon as scheduling information is received, identified as slice type "1", and the other for slices for which the time between reception and processing is not predefined, identified as slice type "2". Although only two slice types are defined, multiple independent instances of each slice type may coexist simultaneously in the system. The following section provides a detailed explanation of both units, highlighting their key differences. 
These units are responsible for assembling all packets to be transmitted per symbol, maintaining slice-specific information, including the slice identification tag, mapped to the \ac{VLAN} and $eAxC\ ID$, and frequency allocation parameters (start \ac{PRB} and number of \acp{PRB}).
The units share symbol buffers, with one buffer per symbol, totaling 14, which are fixed in the 5G New Radio protocol.

The third step is responsible for encapsulating all packets prepared in the previous step. Encapsulation is divided into three stages: (i) application, (ii) transport, and (iii) Ethernet. The encapsulation of the application layer reads the prepared information and assembles the application packet with only the relevant \ac{I/Q} symbols for the current slice. The transport layer encapsulation adds the \ac{eCPRI} header, and the Ethernet encapsulation adds the Ethernet header with the \ac{VLAN} tag attached. In this case, the packet is completed and sent to the Ethernet transceiver. A control unit is defined as the unit that controls everything. This unit communicates with the \ac{O-RU} controller through the fronthaul management plane, configures the \sigla{}, and controls the latch actions.

\subsection*{The pipeline structure}

\begin{figure}[!ht]  
    \begin{center}
        \caption{Simplified pipeline structure.}
        \label{fig:simplePipeline}
        \includegraphics[width=1\textwidth]{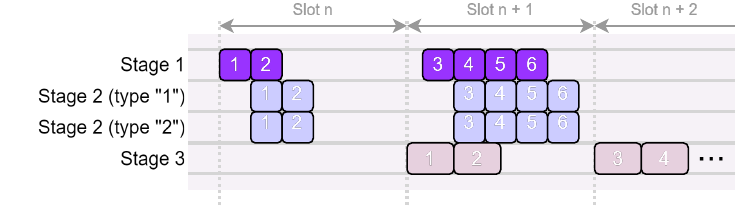}
    \end{center}
\end{figure}

The design of the \sigla{} features a pipeline structure to optimize the utilization of hardware resources on the \ac{FPGA}. This design also reduces processing time by enabling parallelization. Figure \ref{fig:simplePipeline} presents a simplified structure of the adopted pipeline, containing three stages. The first stages represent decoding the C-plane message and processing the scheduling information. The process starts in the first stage when a packet arrives at the \sigla{}. After the information is decoded, the second stage begins, allowing the first stage to become free to receive a new C-plane message simultaneously. As there are two processing units, the second stage runs in parallel. Finally, the third stage occurs based on the time slot change and uses information generated by the second-stage run in the previous time slot. 

We designed the \sigla{} pipeline to be efficient and straightforward. In this case, the three stages have different and variable processing times, resulting in poorly paced behavior at the packet input, as shown in Figure \ref{fig:simplePipeline}, as indicated by the numbered packets. Therefore, buffers are placed between each stage to provide synchronization.

\subsection*{C-Plane Message Decoding}

\ac{O-RU} must receive at least one scheduling message for each existing slice. The C-Plane Message Decoding receives those messages and addresses them to the correct processing unit, depending on the type configured for the respective slice. The control message comprises details about the time and frequency domains for a specific slice. In the time domain, it includes frame, subframe, slot, and symbol information, while in the frequency domain, it specifies the \acp{PRB} utilized on the given symbols.

The $eAxC\ ID$ is extracted from the message as soon as the control message is received. Using this tag, the process consults the control unit to determine whether the corresponding slice is present in the slice list, and sends it to the appropriate processing unit. The control unit maintains the slice list, and the decoding process consults it. If $eAxC\ ID$ is in the control unit list, the decoded data is sent to the type "1" processing unit. Otherwise, the decoded data is sent to the type "2" unit. This design allows a reduced list containing only the slices that must be processed immediately.

\subsection*{Scheduling data processing}

\begin{figure}[!ht]  
    \begin{center}
        \caption{Detail of scheduling data processing.}
        \label{fig:detailSchedProc}
        \includegraphics[width=0.8\textwidth]{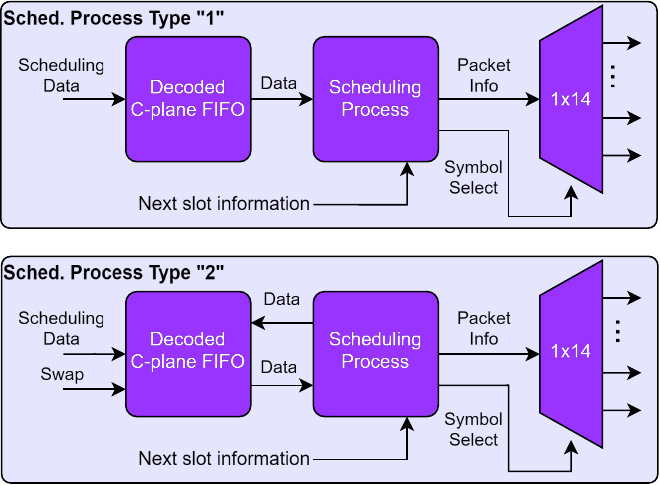}
    \end{center}
\end{figure}

After decoding the C-plane message, all the information is stored in a buffer and read by the processing part. Figure \ref{fig:detailSchedProc} presents the processing unit for considered two types of units. They are very similar but differ in how they handle scheduling information. While the \textit{Type "1"} unit reads the data sequentially, requiring only a simple buffer, the \textit{Type "2"} unit must be able to reinsert data into the C-plane buffer, as the information could be out of order. The processing output consists of the packet parameters and the symbols relevant to this packet. A demultiplexer at the output replicates the data to the appropriate symbol buffers in parallel based on the selection information calculated in the processing part.

To initiate processing, the slot information in the \ac{FIFO} output data must match the slot information received by the \ac{O-RU} synchronization module. This information includes the frame, subframe, and slot identifiers for the next time slot. If this matches the slice information received via control messages, uplink processing will occur in the next time slot.
The key difference in the \textit{Type "2"} processing unit is how it handles scheduling information from the input \ac{FIFO}. If the information is not relevant for the current time slot, it is popped from the \ac{FIFO} and reinserted at the \ac{FIFO} input rather than blocking processing as in the \textit{Type "1"} unit. This approach increases flexibility at the expense of determinism. Depending on the number of slices to be reinserted, this process may take longer and consume more buffer memory.


Scheduling processing prepares the PRB data associated with each slice for transmission in the upcoming NR slot. Each slice is characterized by its allocated PRBs and corresponding symbol positions. Since an NR slot consists of 14 OFDM symbols, 14 dedicated symbol buffers are maintained. The demultiplexer directs the PRBs for each slice to the appropriate symbol buffers based on the scheduling decision. Once organized per symbol, the PRB data must be encapsulated into Ethernet packets for fronthaul transmission. The maximum number of PRBs that can be included in a single packet depends on the network MTU and on the size of each PRB, which in turn is determined by configuration parameters such as numerology and I/Q sample width. As a result, a slice may require multiple Ethernet packets per slot. For each slice, the total number of required packets is computed based on the number of allocated PRBs and the maximum PRB capacity per packet. For example, Slice 1 (S1) contains 61 \acp{PRB}, divided into two packets: one for \acp{PRB} 0 to 30 and another for \acp{PRB} 31 to 60. Additionally, Slice 2 (S2) is shown to have 11 \acp{PRB}, requiring one packet per symbol.

The maximum number of slices depends on the \ac{O-RU} bandwidth and the adopted resource granularity.  For instance, in the FR1 band, with \ac{O-RU} deployment with a bandwidth of 100~MHz (273 \acp{PRB}), and assuming symbol-level slicing granularity, i.e., each \ac{PRB} allocated per OFDM symbol is treated as an independent slice, the theoretical upper bound becomes $273 \times 14 = 3822$ slices per slot.  For $\mu = 1$ (30~kHz subcarrier spacing), slot duration in 5G \ac{NR} is $0.5~\mathrm{ms}$ with 14 symbols.

The proposed architecture was intentionally designed to operate at the highest possible resource granularity, namely at the \ac{PRB}-symbol level. This design choice ensures maximum flexibility, allowing the slicing mechanism to support both control-plane and data-plane transmissions.  The symbol-level implementation provides a structural upper bound and guarantees compatibility with scenarios involving short-duration allocations, including control signaling and future ultra-low-latency configurations, and enables forecasting the feasibility of single-symbol transport blocks for 6G extremely low-latency applications.

\begin{figure}[!ht]  
    \begin{center}
        \caption{Data fragmentation example.}
        \label{fig:fragExample}
        \includegraphics[width=1\textwidth]{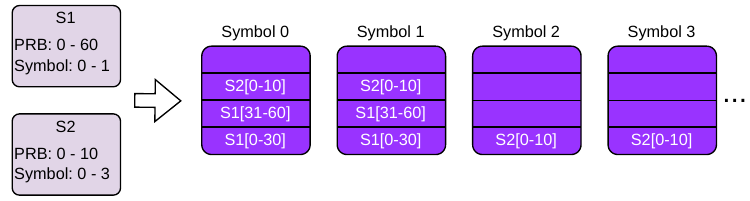}
    \end{center}
\end{figure}

The output signals from each symbol must be combined into a single signal because there are two processing units. This combination is implemented using a 2x1 multiplexer that selects which active processing unit to write to the buffer. The multiplexer prioritizes the \textit{Type "1"} unit over the \textit{Type "2"} unit since the \textit{Type "1"} unit typically has less time to process the scheduling information. If a valid value is available from the \textit{Type "1"} unit, the multiplexer selects it to write to the symbol buffers. The output data from all symbols is combined in a 14x1 demultiplexer, which determines which symbol must be connected to the encapsulation unit. The control unit handles the selection signal and updates the demultiplexer address according to the current symbol.

\subsection*{Encapsulation unit}

\begin{figure}[!h]  
    \begin{center}
        \caption{Encapsulation diagram.}
        \label{fig:flowCap}
        \includegraphics[width=1.0\textwidth]{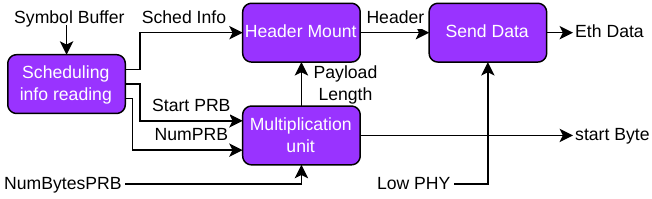}
    \end{center}
\end{figure}

After all packet parameters have been prepared, the encapsulation unit assembles the fronthaul packets. The encapsulation process comprises several steps, described in Figure \ref{fig:flowCap}. 

The first step is to read the packet parameters from the symbol buffer. If an entry is available in the buffer output, the encapsulation unit reads it. Therefore, the payload size is calculated with the loaded information. This operation is performed in parallel by an arithmetic unit and follows Equation \ref{eq: payloadSize}, where $IQ_{w}$ is the size of \ac{I/Q} symbols in bits, $N_{PRB}$ is the number of \acp{PRB}, and $O_{App}$ is the application overhead, which is equal to 12, as specified by the \ac{O-RAN} Alliance \cite{oran.cusFront.wg4.2023}). At the same time, the first byte in the LOW-\ac{PHY} is calculated, which informs the LOW-\ac{PHY} processing unit of which byte the data must be sent from. Similarly to the payload size calculation, this operation is performed by a separate multiplication unit, following Equation \ref{eq: startByte}, where $IQ_{w}$ is the size of \ac{I/Q} symbols in bits, and $S_{PRB}$ is the starting \ac{PRB}.

\begin{equation}
    \label{eq: payloadSize}
    PL = (3 \cdot IQ_{w} \cdot N_{PRB}) + O_{App},
\end{equation}

\begin{equation}
    \label{eq: startByte}
    SB = 3 \cdot IQ_{w} \cdot S_{PRB}.
\end{equation}

The packet information stored in the symbol buffers is not sorted by \acp{PRB}, making communication with the LOW-\ac{PHY} in a single-stream run impossible. Moreover, storing LOW-\ac{PHY} data would require significant memory resources and processing time. Therefore, a start-byte signal was added to the interface. The LOW-\ac{PHY} unit is expected to have access to the data across all bandwidths, allowing it to prepare the stream upon receiving a start byte.
The encapsulation process starts as soon as the payload length is calculated. The application, transport, and data link headers are filled and sent to the Ethernet transceiver. It is expected that by the time the header has been sent, the LOW-\ac{PHY} has already configured the interface with the appropriate first byte. The process continues until the sent data reaches the previously calculated payload length, completing the packet encapsulation. These steps are repeated until the symbol buffer becomes empty, completing the encapsulation of that symbol. For the remainder of the time, the unit waits for the symbol to change and for the symbol buffer to be updated, then restarts the entire process.

\subsection*{Control unit}

\begin{figure}[!ht]  
    \begin{center}
        \caption{Control Unit block diagram.}
        \label{fig:blockDiaCUnit}
        \includegraphics[width=0.8\textwidth]{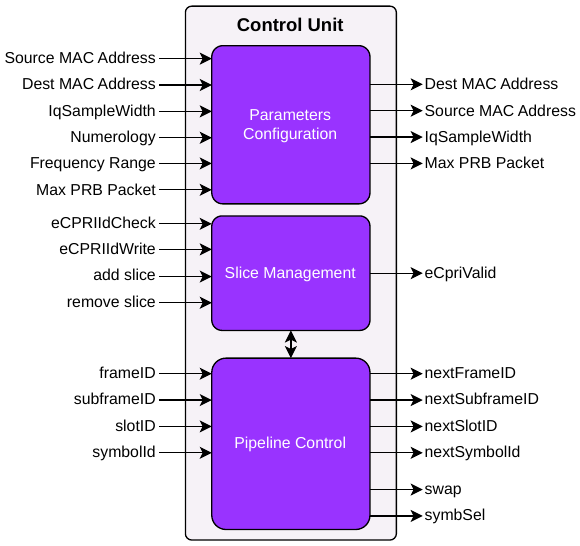}
    \end{center}
\end{figure}

The control unit is responsible for ensuring that all other blocks work together. This unit handles some functions to do that, as presented in Figure \ref{fig:blockDiaCUnit}. The pipeline control ensures that all other blocks in the proposed \sigla{} work together, to generate the correct signals at the right time. Slice management handles slice creation, maintenance, edition, and destruction. The parameter configuration, such as \ac{MAC} addresses, takes some general parameters.

\section{Performance evaluation}
\label{sec:metrics}

This section presents the scenarios designed to test and validate the performance of the \sigla{}. First, we describe the prototype, followed by the metrics used to evaluate it. Next, we present two use cases, and we discuss the results.

\subsection*{\textbf{\sigla{} prototype}}

The prototype project aims to create an experimental environment for the \sigla{}, focusing on the Receiver pipeline scheme to meet the New Radio timing requirements. In this context, we use the KCU105 platform to develop the \sigla{}. This platform is a development kit for \ac{AMD} Xilinx's Kintex UltraScale \ac{FPGA}. This high-end \ac{AMD} product is well-suited for applications that require parallelism, low latency, power efficiency, and real-time processing, making it an ideal component for developing the \sigla{}.

\begin{figure}[!ht]  
    \begin{center}
        \caption{\sigla{} prototype.}
        \label{fig:proto}
        \includegraphics[width=1\textwidth]{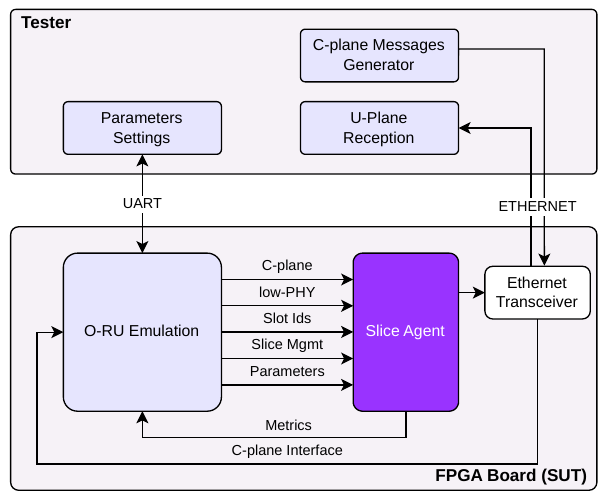}
    \end{center}
\end{figure}

Figure \ref{fig:proto} shows the prototype architecture. We developed the \sigla{} on the \ac{FPGA} board, receiving control messages and sending uplink user-plane data via the Ethernet interface. Currently, no functional \ac{O-RU} project has been identified to serve as the foundation for this project, so we have also developed an \ac{O-RU} emulation block to generate time-reference signals and overall parameters. Moreover, we developed a Python application to simulate the management plane and \acp{O-DU}. This software is responsible for generating C-plane messages over the Ethernet interface and for configuring the \sigla{} with the correct parameters via a \ac{UART} interface. Furthermore, the metrics collected during emulation are read by the software over \ac{UART} to evaluate the \sigla{}'s performance. The received user uplink data is captured using Wireshark, which already supports \ac{eCPRI} protocol formatting, facilitating data analysis.

We can set up all parameters for the \ac{O-RU} emulation through the software component called \textit{Tester}, which we wrote in Python. The \textit{Tester} can generate configuration parameters, including numerology, bandwidth, the width of the \ac{I/Q} samples, and slice configuration, for the \ac{SUT}. A variable number of slices with specific characteristics can be configured in the slice configuration, including the time and frequency domains of uplink data and the timing of when \sigla{} receives these messages. This point is essential for emulating various fronthaul conditions, including jitter and latency configurations for different slice types. The source code for the \sigla{} prototype, \textit{Test Cases}, and additional evaluation tools developed are available on GitHub\footnote{https://github.com/cbboth/slice-agent-shared-oru/}. The \ac{O-RU} emulation block is explained in more detail in the next section.

\subsection*{\ac{SUT} sequence}

\begin{figure}[!ht]  
    \begin{center}
        \caption{\ac{SUT} sequence diagram.}
        \label{fig:protoTime}
        \includegraphics[width=1\textwidth]{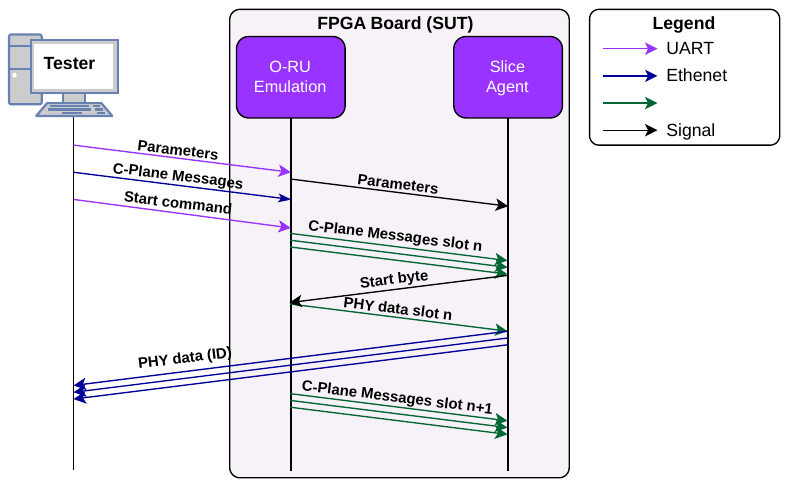}
    \end{center}
\end{figure}

The testing process must follow a specific order to ensure everything functions correctly. Figure \ref{fig:protoTime} presents a basic sequence diagram illustrating this process. Initially, the \textit{Test Cases} sends all necessary parameters to the \ac{FPGA} board over \ac{UART}, i.e., to the \ac{SUT}. These parameters include numerology, the width of the I/Q samples, the number of \acp{PRB} per Ethernet packet, symbol period, and initial time slot identifications. The start clock cycle is also sent for each subsequent C-plane message. These details are stored in a \ac{FIFO} and must be sent in the order of their occurrence to prevent the emulation from blocking. The \ac{O-RU} emulation block forwards the relevant parameters to the \sigla{}.

All C-plane messages are transmitted over Ethernet to the \ac{FPGA} board. These messages are also stored in a \ac{FIFO} and must align with the start clock cycle order sent over \ac{UART}. Once the start command is issued, the \ac{O-RU} emulation begins sending C-plane messages to the \sigla{} via an AXI4-Stream interface, synchronized with the clock counter and starts clock cycle. The \sigla{} processes these messages, and when encapsulation begins, it signals the start byte of the \ac{PHY} data. Upon receiving this signal, the \ac{O-RU} emulator initiates the transmission of \ac{PHY} data over the AXI4-Stream interface.
The \sigla{} then sends the encapsulated \ac{PHY} data, with the correct slice identification, to the \textit{Test Cases}. This process continues until all C-plane messages have been processed.

\subsection*{\textbf{Metrics}}

We employed several metrics in the analysis to understand the performance and resource utilization of the \sigla{}. These metrics provide a comprehensive view of the \sigla{}'s efficiency, processing capabilities, and resource demands. The three primary metrics used were:

\begin{itemize}
    \item \textit{Processing time}: measured in clock cycles, processing time indicates how long it takes for the \sigla{} to handle specific tasks. This metric includes the time from receiving scheduling information to the encapsulation and transmission of data. Processing time is a crucial metric because it directly affects the system's latency and throughput.
    \item \textit{Memory occupation}: evaluates the amount of memory the \sigla{} uses, particularly focusing on the buffers and \acp{FIFO} used in the processing units. Efficient memory usage is crucial to ensure that \ac{FPGA} resources are utilized optimally without causing bottlenecks or resource contention.
    \item \textit{Resource utilization}: includes the usage of \acp{LUT}, \acp{FF}, \ac{BRAM}, and \ac{DSP} blocks in the \ac{FPGA}. High resource utilization can indicate potential areas for optimization, while low utilization suggests efficient design.
\end{itemize}

\subsection*{\textbf{Use cases}}

We validated the \sigla{} by testing two primary use cases: (i) \ac{mMTC} and (ii) \ac{URLLC}. In this case, we designed each scenario to evaluate the \sigla{}'s performance and resource utilization under different conditions. The following subsections detail the specific parameters and configurations used in these scenarios.

\subsubsection*{\ac{mMTC} use case}
\label{sec: mmtccase}

The \ac{mMTC} scenario simulates a high-density network environment where numerous small devices transmit data packets. This use case focuses on handling many slices with relatively small \acp{PRB}. The configurations tested include:
\begin{itemize}
    \item \textit{Medium-density \ac{mMTC}}: 600 slices of 1 \ac{PRB} each were created in this scenario. The number of slices demonstrates a small chance that the \sigla{} \ac{FIFO} is full.
    \item \textit{High-density \ac{mMTC}}: 1200 slices of 1 \ac{PRB}, i.e., the number of slices was doubled. This number is higher than the \ac{FIFO} implemented, demonstrating the failure of the \sigla{}.
\end{itemize}

For both scenarios, we randomized the reception times of C-plane messages to simulate different slice requirements.

\subsubsection*{URLLC use case}
\label{sec: urllccase}

The \ac{URLLC} scenario focuses on low-latency communication for critical applications. Each \ac{URLLC} slice consists of more \acp{PRB} and requires immediate processing to meet stringent latency requirements. The \sigla{}'s ability to prioritize and handle these urgent messages without delay is crucial for maintaining service quality. The \ac{URLLC} use case was tested alongside \ac{mMTC} to evaluate the \sigla{}'s isolation ability. We consider two scenarios:

\begin{itemize}
    \item \textit{\ac{URLLC} slices marked as type "2"}: we created four slices with 30 \acp{PRB} each. In this scenario, the processing unit is shared with the \ac{mMTC} slices.
    \item \textit{\ac{URLLC} slices marked as type "1"}: The same configuration as the previous scenario, but in this case, the \ac{URLLC} slices are processed in a different unit.
\end{itemize}

Although the experimental validation focuses on these two use cases, the proposed architecture is not restricted to them. The selected scenarios were intentionally chosen to stress complementary system dimensions: scalability under massive connectivity (\ac{mMTC}) and strict latency requirements (\ac{URLLC}). The same architectural principles extend naturally to other service categories, including \ac{eMBB}. From the perspective of the proposed slicing mechanism, an \ac{eMBB} scenario would be structurally similar to \ac{mMTC}, with the primary difference being the larger number of \acp{PRB} allocated per slice to accommodate broadband traffic demands. The evaluations encompassing \ac{mMTC} and \ac{URLLC} use cases demonstrate the \sigla{}'s ability to handle high-density traffic and low-latency demands, respectively. The subsequent section provides a detailed analysis of the results and performance, offering a deeper understanding of the \sigla{}'s capabilities and limitations.

\subsection*{\textbf{Evaluations}}

This section discusses the results in two groups: performance analysis and hardware resources analysis. In the first group, the evaluations aimed to measure the effectiveness and efficiency of the \sigla{} under different conditions and workloads. The second group presents a discussion of the hardware resources based on an \ac{FPGA} device to develop the \sigla{}.

\subsubsection*{Performance analysis}

The scheduling data processing is the main parameter that determines how long before the first uplink data the reception window must open. This step consists of decoding the information from the previous step and preparing all the uplink data packets sent during the encapsulation phase.
The time required to process a single slice depends on the clock speed, the size of a slice in terms of \acp{PRB}, the Ethernet \ac{MTU}, and the width of the \ac{I/Q} samples. Processing starts when the decoded C-plane data is read from the \ac{FIFO} and finishes when the packet data is written to the symbol \ac{FIFO}. 

\begin{figure}[!ht]  
    \begin{center}
        \caption{Processing time of a slice based on number of \acp{PRB} and \ac{MTU}.}
        \label{figR:tprocVsPrb}
        \includegraphics[width=1\textwidth]{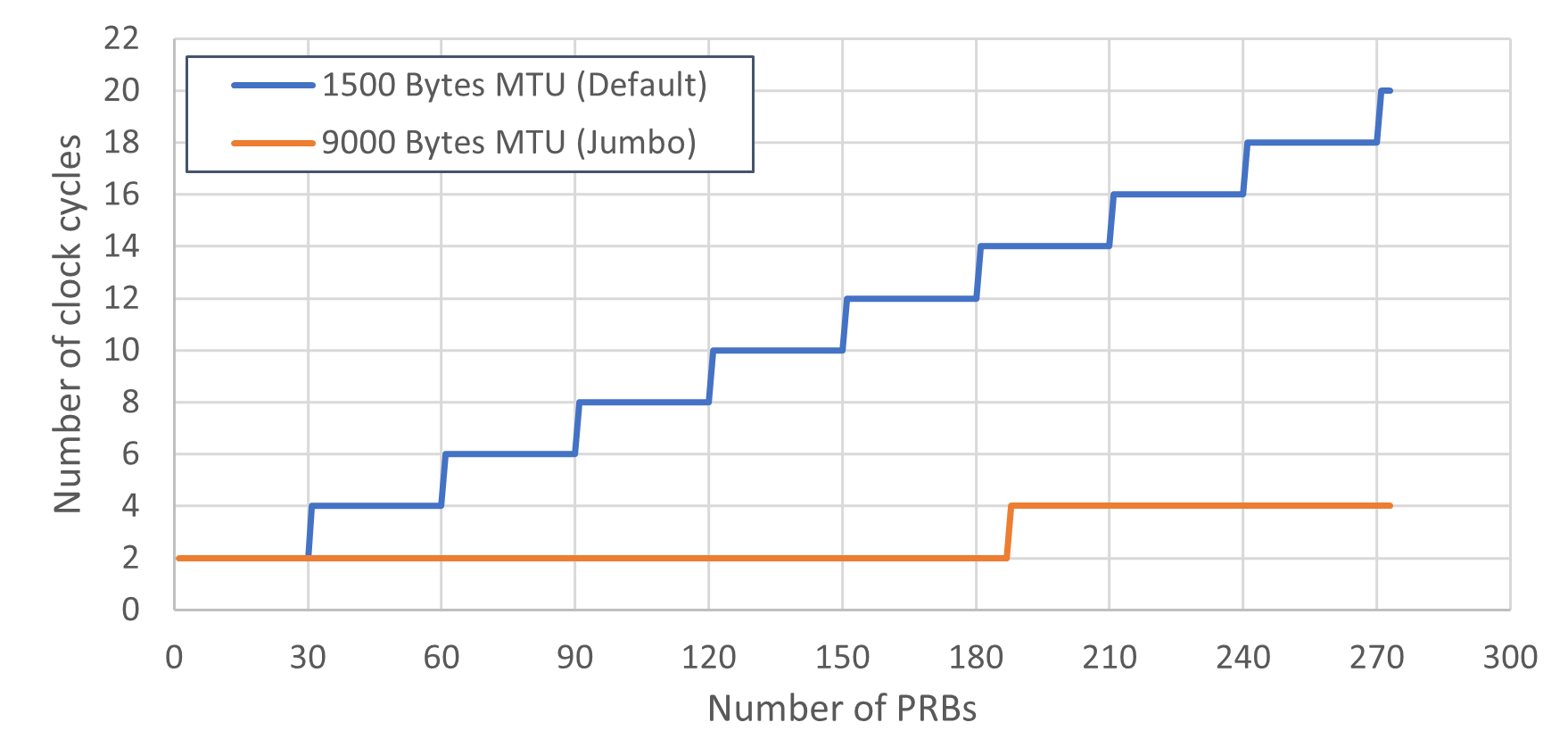}
    \end{center}
\end{figure}

Figure \ref{figR:tprocVsPrb} presents the performance of the implemented design, considering that the width of the \ac{I/Q} samples is 16 bits, which is the uncompressed value. Processing time is directly related to the number of packets needed for a specific slice. For one packet, the processing time is 2 clock cycles, and this time increases linearly with the number of packets. For an \ac{MTU} of 1500 bytes, the maximum default value of the Ethernet protocol, the maximum number of \acp{PRB} is 30 within the same packet. This maximum value is due to the processing time increasing by two clock cycles every 30 \acp{PRB}. For Jumbo packets, the \ac{MTU} is increased to 9000 bytes, and for every slice with less than 188 \acp{PRB}, just one packet is needed, and the processing time is two clock cycles. Two packets are required for this case, increasing the processing time to four clock cycles.

The total processing time depends not only on the size of the slice in terms of \acp{PRB}, but also on the number of slices. Equations (\ref{eq: procTime}) and (\ref{eq: nPkt}) determine the processing time, considering the parameters explained in Table \ref{tab:tProcParams}. For every packet, an additional clock cycle is needed to write the processed information to the buffer, and another to start processing within a time slot.

\begin{equation}
    \label{eq: procTime}
    t_{proc}[clk\ cycles] = 1 + \sum_{s=1}^{N_s} (2 \cdot n_{pkt}^{s}) + 1
\end{equation}
\begin{equation}
    \label{eq: nPkt}
    n_{pkt}^{s} = ceil\left(\frac{n_{PRB}^{s}}{n_{PRBpkt}}\right)
\end{equation}

\begin{table}[!ht]
\caption{Parameters definition for processing time equation.}
\label{tab:tProcParams}
\centering
\begin{tabular}{l c}
\hline \hline
     \textbf{Variable}      & \textbf{Meaning}\\ \hline \hline
     \rowcolor{Gray}$N_s$   &   Number of slices \\ \hline
     $n_{pkt}^{s}$          &   Number of packet for slice $s$  \\  \hline
     \rowcolor{Gray} $n_{PRB}^{s}$  &   Number of \acp{PRB} for slice $s$    \\  \hline
     $n_{PRBpkt}$           &   Maximum number of \acp{PRB} in an Ethernet packet  \\ \hline
    \hline
\end{tabular}
\end{table}

A slice is determined in the frequency domain by the number of \acp{PRB} and in the time domain by the number of symbols. In our design of the \sigla{}, this characterization plays a crucial role in how slices are processed and scheduled. As demonstrated, the processing time of the \sigla{} depends on the number of necessary packets rather than the number of symbols. This time is because the processed information is written simultaneously to the relevant symbols in parallel. In this context, parallel writing ensures that increasing the number of symbols does not adversely affect the processing time. In this case, our design gives the scheduler flexibility in resource allocation. Moreover, it can choose to place a slice within a single symbol containing a high number of \acp{PRB} or distribute the \acp{PRB} across multiple symbols, resulting in fewer \acp{PRB} per symbol but more symbols overall.

\begin{figure}[!ht]  
    \begin{center}
        \caption{Number of packets and processing time for one slice of (a) 30, (b) 90, (c) 210, (d) 273 \acp{PRB}.}
        \label{figR:tProcVsSymb}
        \includegraphics[width=1\textwidth]{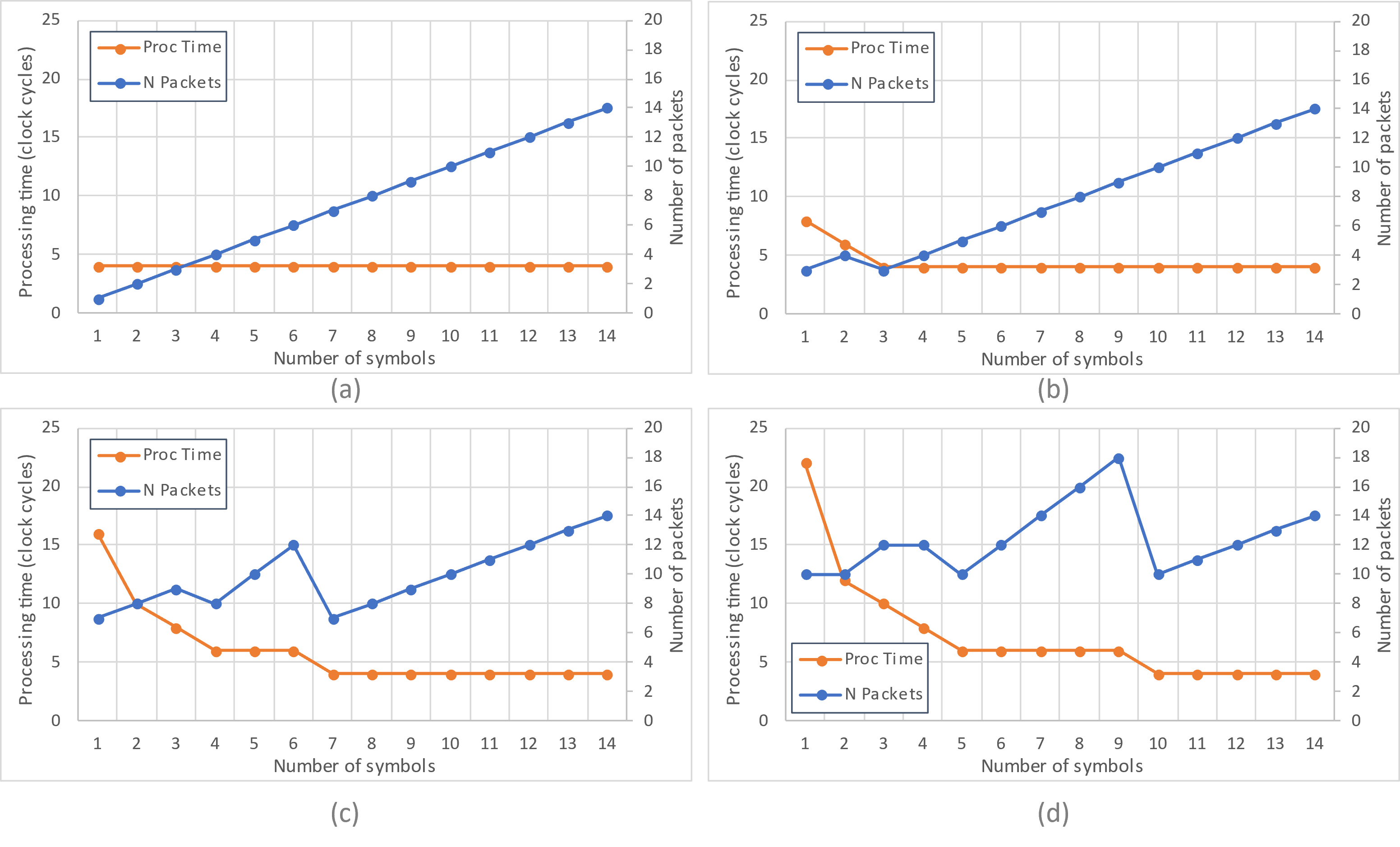}
    \end{center}
\end{figure}

It is fundamental to analyze the optimal division that minimizes both processing time and the total number of packets, considering a fixed total number of \acp{PRB}. This division changes approximately every 30 \acp{PRB} with the default \ac{MTU} of 1500 bytes. Figure \ref{figR:tProcVsSymb} presents the division for four scenarios: 30, 90, 210, and 273 \acp{PRB}. It is evident that for 30 \acp{PRB}, maintaining all of them within one symbol yields the best result. Considering 90 \acp{PRB}, the optimal division rises to three symbols, 210 \acp{PRB} to seven symbols, and 273 \acp{PRB} to 10 symbols. In this analysis, we can observe that dividing 10 slices of 273 full-bandwidth \acp{PRB} into 10 symbols each is more efficient than keeping all PRBs in a single symbol per slice. In this case, we reduce the processing time by a factor of 10 while maintaining the same number of packets.

The scheduler can reduce the overhead associated with packet processing and transmission by balancing the number of \acp{PRB} and symbols. Finding this optimal balance is essential for efficient system performance, ensuring that resources are utilized effectively without unnecessary delays.
In the context of C-plane message processing, two processing units were defined: (i) one for critical slices and (ii) another for general slices, referred to as \textit{Type "1"} and \textit{Type "2"} units. The processing time for both units is identical. However, since the \textit{Type "2"} unit can return data to the \ac{FIFO}, its processing time also depends on the number of slices already in the \ac{FIFO} from previous time slots. If the number of elements reaches the maximum supported by the \ac{FIFO}, any additional C-plane messages are discarded. However, the \textit{Type "1"} processing unit does not face this issue but requires that C-plane messages be received in order. Otherwise, the unit will lock, affecting the radio unit's functionality.

\begin{figure}[!ht]  
    \begin{center}
        \caption{\textit{Type "2"} processing unit input \ac{FIFO} occupation for 600 and 1200 slices per slot.}
        \label{figR:ocu1}
        \includegraphics[width=1\textwidth]{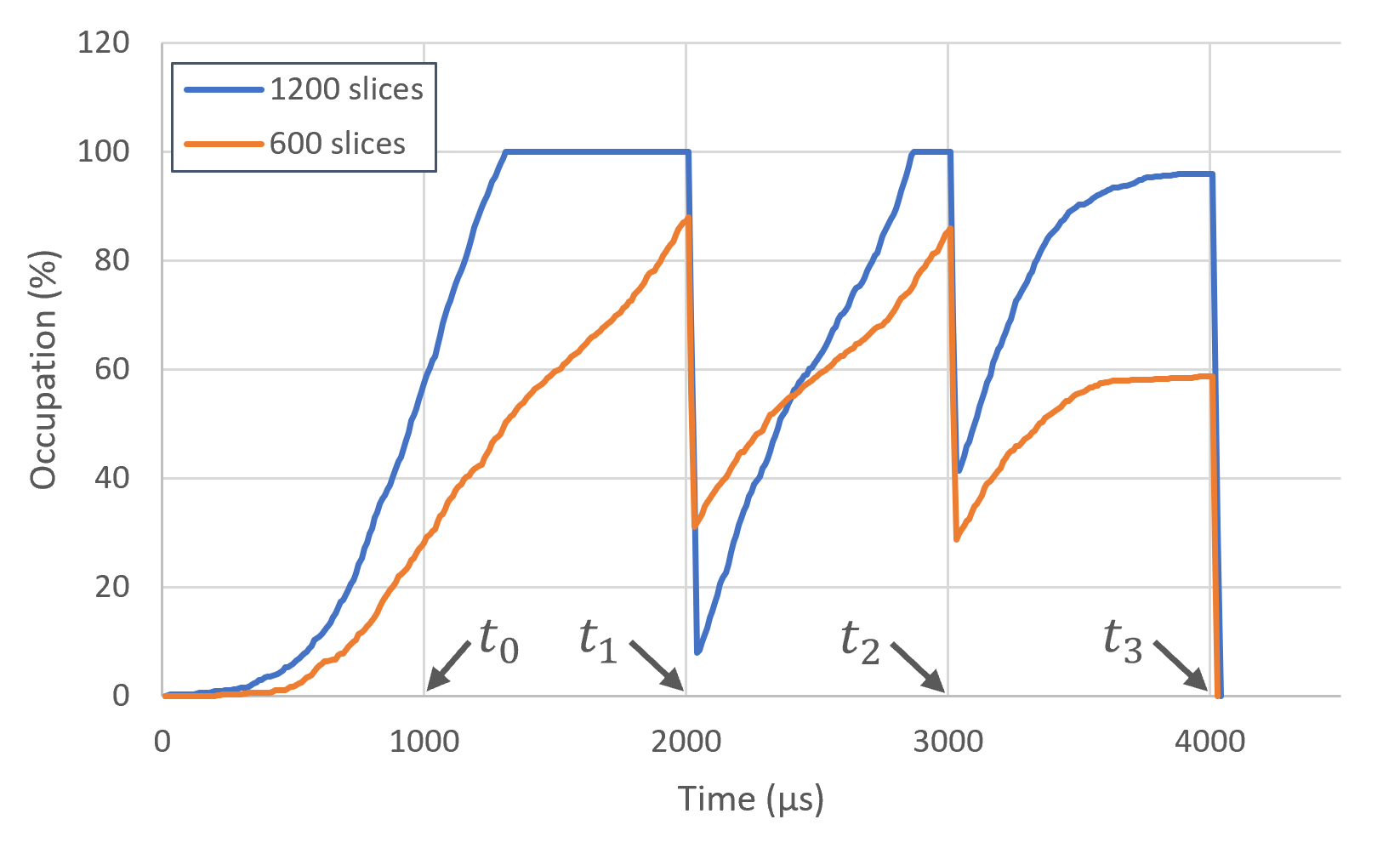}
    \end{center}
\end{figure}

We used the \ac{mMTC} use case to evaluate the constraints of the \textit{Type "2"} unit. The \ac{FIFO} can hold up to 1024 elements, and to simulate a real scenario, we randomize the reception timing of C-plane messages that occur within the previous two time slots. This experiment results in messages being received out of order. In the first scenario, we used 600 slices per time slot, which is about half of the \ac{FIFO}'s capacity. However, the out-of-order messages can still cause a \ac{FIFO} overflow. In the second scenario, we increased the number of slices to 1200, exceeding the \ac{FIFO} capacity. Figure~\ref{figR:ocu1} shows the \ac{FIFO} occupation for both scenarios. We can observe that the \ac{FIFO} overflowed with 1200 slices, causing C-plane message loss. In this experiment, three consecutive time slots were represented, totaling 3600 slices. Out of these 3600 slices, 981 were lost, i.e., 27.25\%. The processing time is very short, so with each time slot swap, the \ac{FIFO} occupation decreases rapidly, leaving only the slices that will be processed in future time slots.

Finally, it is essential to note that the processing units are isolated from each other, allowing critical slices to be assigned to the \textit{Type "1"} unit. We evaluated this isolation using the \ac{URLLC} use case. In the first scenario, where \ac{URLLC} and \ac{mMTC} share the \textit{Type "2"} unit, all \ac{URLLC} slices were lost because, by the time they were received, the \ac{FIFO} was already full. This behavior did not occur in the second scenario, where the \ac{URLLC} slices were configured to be processed by the \textit{Type "1"} unit. Even when the \textit{Type "2"} unit's \ac{FIFO} overflowed, the \textit{Type "1"} unit remained unaffected, processing all \ac{URLLC} slices without issues. This setup ensures isolation between the two units, making the \sigla{} suitable for various use cases.


\subsubsection*{Hardware resources analysis}

We synthesized and implemented the \sigla{} in a Kintex 7 Ultrascale \ac{AMD} \ac{FPGA}, model XCKU040-FFVA1156-2-E. Table \ref{tab:saResources} presents the resources used, showing the quantity and percent usage for the specific \ac{FPGA}. From the list, two resources show a significant amount: (i) the number of \ac{LUT} and (ii) the quantity of \ac{BRAM} as \ac{FIFO}, i.e., more specifically FIFO36E2. 
Considering the 35 \acp{FIFO} used, 28 are for the symbol buffers, i.e., 2 per symbol. In this case, it is possible to customize the size to meet a specific application requirement, which can impact the number of resources used. For sizes greater than 512, the number of memories used grows linearly with size. If 1024 is selected, four FIFO36E2 are used for each symbol buffer. However, for sizes smaller than 512, the number of resources remains constant since a \ac{FIFO} cannot be shared in the \ac{FPGA} used.

\begin{table}[!ht]
\caption{\sigla{} resource utilization.}
\label{tab:saResources}
\centering
\begin{tabular}{l c c}
\hline \hline
     \textbf{Resource type}                 & \textbf{Used}     &   \textbf{Utilization (\%)} \\ \hline \hline
     \rowcolor{Gray} LUT as Logic              &   4555            &   1.88   \\ \hline
     Register as Flip Flop     &   1909            &   0.39   \\ \hline
     \rowcolor{Gray}FIFO36E2                  &   35              &   5.83   \\ \hline
     DSP48E2                   &   3               &   0.16   \\ \hline
    \hline
\end{tabular}
\end{table}

From the perspective of \ac{LUT} and Flip Flop, only the list that stores the configured slices consumes 1136 \acp{LUT} and 556 Flip Flops, corresponding to $48.94\%$ of \acp{LUT} and $29.12\%$ of Flip Flops used by the \sigla{}, as can be seen in Table \ref{tab:listSliceRes}. This consumption occurs because implementing this list requires that all positions be accessible simultaneously, making the use of \ac{BRAM}, as in symbol buffers, impossible. For these values, the size of this list is 32, limiting the maximum number of slices of \textit{Type "1"}. Although this number is small compared to the total possible slices (3822), the total number of slices of \textit{Type "1"} (usually \ac{URLLC}) is highly dependent on the application. Increasing the list size allows more slices to be marked as \textit{Type "1"} at the cost of increased hardware resources. Table \ref{tab:listSliceRes} also presents the resources used for a list with sizes 256, 128, 64, and 32.

\begin{table}[!ht]
\caption{Resources of the list of configured slices.}
\label{tab:listSliceRes}
\centering
\begin{tabular}{l c c c c}
\hline \hline
     \multirow{2}{*}{\textbf{Resource type}} & \multicolumn{4}{c}{\textbf{List size}} \\
                                        & \textbf{256} & \textbf{128} & \textbf{64} & \textbf{32}  \\ \hline \hline

    \rowcolor{Gray} LUT                    &   9718    & 4955  &  2176 & 1136  \\ \hline
    Register as Flip Flop  &   5155    & 2648  &  1088 & 556   \\ \hline
    \hline
\end{tabular}
\end{table}

The amount of resources used shows an almost linear relationship with the list size. For a list of size 256, the list uses more than 5\% of the available \acp{LUT}. While this utilization may not seem excessive, given that the \sigla{} is just a small part of the \ac{O-RU} context, it could limit future implementations. Therefore, in a use case where only a few slices are needed, reducing the list's size saves significant resources. 
Lastly, the usage of the DSP48E2 resource enables low-latency multiplication to calculate packet length in bytes. This procedure is performed during encapsulation, and doing it as consecutive sums could increase encapsulation time, waste time, and possibly fail to meet data transmission requirements.

\section{Conclusion}
\label{sec:conc}

\ac{5G} and forthcoming \ac{6G} systems introduce a paradigm shift in cellular network design, driven by virtualization, disaggregation, and intelligence at the \ac{RAN} level. Within this context, \ac{O-RAN} plays a central role by enabling open interfaces and programmable components that support advanced functionalities such as Network Slicing. While slicing is conceptually defined at higher layers, its practical enforcement at the fronthaul level introduces non-trivial architectural and hardware challenges, particularly within the \ac{O-RU}, where cost, latency, and processing constraints are critical.

This work proposed and implemented \sigla{}, a hardware-based slicing support mechanism for the \ac{O-RU} fronthaul. The proposed architecture enables real-time slice identification and segregation at the packet-processing stage, supporting a large number of heterogeneous slices. The prototype demonstrated deterministic processing with a latency of 2 clock cycles per packet, independent of the slice configuration.

The theoretical analysis showed that, for a bandwidth of 100~MHz, up to 3822 slices may coexist within a single slot when each \ac{PRB}-symbol tuple is treated as an independent slice. This scenario stresses memory indexing and lookup structures and validates the scalability limits of the proposed design. Experimental evaluation under \ac{mMTC} and \ac{URLLC} configurations confirmed the system's capability to handle both high slice density and stringent latency constraints. In particular, memory-occupancy analysis revealed that more than 5~\% of the available \ac{BRAM} resources were required for slice management, while parallel list structures consumed up to 48.94~\% of the \acp{LUT} resources in larger configurations. These results highlight the inherent trade-off between scalability and hardware cost in disaggregated \ac{RAN} architectures.

From a practical deployment perspective, the results indicate that implementing slicing enforcement at the \ac{O-RU} level is feasible but requires careful dimensioning of memory structures and parallel processing units. The design choices directly impact \ac{FPGA} resource utilization and may constrain the integration of additional \ac{O-RU} functionalities. Therefore, slicing support in disaggregated architectures must be co-designed with hardware constraints rather than treated as a purely logical function.

Future work includes integrating with the \ac{SMO} to expose real-time metrics (e.g., \ac{FIFO} occupancy and slice processing latency), enabling closed-loop optimization via \ac{RIC}-based control. Further validation in network simulators (e.g., NS-3) and integration with commercial \acp{O-RU} will allow assessment of the \sigla{} under realistic traffic and deployment conditions.

\section*{Acknowledgment}

This work was conducted with partial financial support from the National Council for Scientific and Technological Development (CNPq) under grant number 308720/2025-3 and MCTIC/CGI.br/São Paulo Research Foundation through project Programmability, ORchestration and VIRtualization of 5G Networks (PORVIR-5G) n$^o$ 2020/05182-3.

\bibliographystyle{IEEEtran}
\bibliography{references.bib}

@online{gsma.5gGuide.2019,
    title = {{The 5G Guide: A Reference for Operators}},
    author = {{GSM Association}},
    year = {2019},
    url = {https://www.gsma.com/wp-content/uploads/2019/04/The-5G-Guide_GSMA_2019_04_29_compressed.pdf},
    note = {{Accessed on August 21, 2023}}
}

@techreport{oran.oad.wg1.2023,
    author = {{O-RAN Alliance}},
    number = {{Work Group 1 - Version 9}},
    title = {{O-RAN Architecture Description 9.0}},
    type = {{TS}},
    institution = {{Open Radio Access Network Alliance}},
    url = {https://orandownloadsweb.azurewebsites.net/specifications},
    year = {2023},
    note = {{Accessed on August 21, 2023}}
}

@techreport{oran.cusFront.wg4.2023,
    author = {{O-RAN Alliance}},
    number = {{Work Group 4 - Version 12}},
    title = {{Control, User and Synchronization Plane Specification}},
    type = {{TS}},
    institution = {{Open Radio Access Network Alliance}},
    url = {https://orandownloadsweb.azurewebsites.net/specifications},
    year = {2023},
    note = {{Accessed on August 21, 2023}}
}

@techreport{oran.mFront.wg4.2023,
 author = {{O-RAN Work Group 4}},
 institution = {{Open Radio Access Network Alliance}},
 month = {06},
 note = {Version 12},
 title = {{Management Plane Specification}},
 type = {{TS}},
 url = {https://orandownloadsweb.azurewebsites.net/specifications},
 year = {2023}
}

@techreport{oran.usecases.wg1.2023,
    author = {{O-RAN Alliance}},
    number = {{Work Group 1 - Version 10}},
    title = {{Use Cases Detailed Specification}},
    type = {{TS}},
    institution = {{Open Radio Access Network Alliance}},
    url = {https://orandownloadsweb.azurewebsites.net/specifications},
    year = {2023},
    note = {{Accessed on August 21, 2023}}
}

@techreport{oran.slice.wg1.2023,
    author = {{O-RAN Alliance}},
    number = {{Work Group 1 - Version 10}},
    title = {{Slicing Architecture}},
    type = {{TS}},
    institution = {{Open Radio Access Network Alliance}},
    url = {https://orandownloadsweb.azurewebsites.net/specifications},
    year = {2023},
    note = {{Accessed on August 21, 2023}}
}

@techreport{oran_wg1_slicing_v14_2025,
  title        = {{O-RAN Slicing Architecture}},
  institution  = {O-RAN Alliance},
  author       = {{O-RAN Alliance WG1}},
  type         = {Technical Specification},
  number       = {O-RAN.WG1.TS.Slicing-Architecture-R004-v14.01},
  year         = {2025},
  url          = {https://specifications.o-ran.org},
  note         = {Version 14.01}
}

@techreport{oran_wg4_cus_v20,
  title        = {{O-RAN Control, User and Synchronization Plane Specification}},
  institution  = {O-RAN Alliance WG4},
  number       = {O-RAN.WG4.TS.CUS.0-R005-v20.00},
  year         = {2026},
  url          = {https://specifications.o-ran.org/download?id=738},
  note         = {Open Fronthaul interface specification. Accessed: 2026-04-23}
}

@misc{3gpp20173gpp,
  title={{3GPP ts 23.501 technical specification group services and system aspects; system architecture for the 5G system; stage 2 (release 15)}},
  author={3GPP},
  year={2017}
}

@techreport{3gpp.28.530,
 author = {3GPP},
 month = {09},
 note = {Version 17.3.0},
 number = {38.801},
 title = {{Technical Specification Group Services and System Aspects; Management and orchestration;
Concepts, use cases and requirements 

}},
institution = {3GPP},
 year = {2022}
}

@techreport{3gpp_ts23501_r20,
  author       = {{3GPP}},
  title        = {{TS 23.501: System Architecture for the 5G System (5GS)}},
  institution  = {3rd Generation Partnership Project (3GPP)},
  type         = {Technical Specification},
  number       = {23.501},
  version      = {20.0.0},
  year         = {2025},
  month        = dec,
  note         = {Release 20, Stage 2}
}

@techreport{3gpp_ts23502_r20,
  author       = {{3GPP}},
  title        = {{TS 23.502: Procedures for the 5G System (5GS)}},
  institution  = {3rd Generation Partnership Project (3GPP)},
  type         = {Technical Specification},
  number       = {23.502},
  version      = {20.0.0},
  year         = {2025},
  month        = dec,
  note         = {Release 20, Stage 2}
}

@techreport{3gpp_ts38401_r19,
  author       = {{3GPP}},
  title        = {{TS 38.401: NG-RAN; Architecture description}},
  institution  = {3rd Generation Partnership Project (3GPP)},
  type         = {Technical Specification},
  number       = {38.401},
  version      = {19.1.0},
  year         = {2025},
  month        = dec,
  note         = {Release 19}
}

@techreport{3gpp.38.300.2018,
    author = {3GPP},
    number = {Version 15.0.0},
    title = {{NR; NR and NG-RAN Overall description; Stage-2}},
    type = {{TS}},
    institution = {{3rd Generation Partnership Project}},
    url = {https://portal.3gpp.org/desktopmodules/Specifications/SpecificationDetails.aspx?specificationId=3191},
    year = {2018}
}

@techreport{3gpp.38.300.r16.2023,
    author = {3GPP},
    note = {Version 16.12.0},
    number = {38.300},
    title = {{NR; NR and NG-RAN Overall description; Stage-2}},
    type = {{TS}},
    institution = {{3rd Generation Partnership Project}},
    url = {https://portal.3gpp.org/desktopmodules/Specifications/SpecificationDetails.aspx?specificationId=3191},
    year = {2023}
}

@techreport{3gpp.38.300.r17.2023,
    author = {3GPP},
    note = {Version 17.4.0},
    number = {38.300},
    title = {{NR; NR and NG-RAN Overall description; Stage-2}},
    type = {{TS}},
    institution = {{3rd Generation Partnership Project}},
    url = {https://portal.3gpp.org/desktopmodules/Specifications/SpecificationDetails.aspx?specificationId=3191},
    year = {2023}
}

@techreport{ITU-G-Sup71-2023,
  author = {{ITU-T}},
  title        = {{Supplement 71 to ITU-T G-series Recommendations: Optical line termination capabilities for supporting cooperative dynamic bandwidth assignment}},
  type         = {TR Supplement},
  number       = {G Suppl. 71},
  year         = {2023},
  address      = {Geneva, Switzerland},
  edition      = {2.0},
  url          = {https://handle.itu.int/11.1002/1000/15854}
}

@techreport{etsi.zsm.2021,
    author = {{ETSI GS ZSM}},
     year = {2021},
    note = {Version 1.1.1},
    title = {{Zero-touch network and Service Management (ZSM); End-to-end management and orchestration of network slicing}},
    type = {{TS}},
    institution = {{European Telecommunications Standards Institute}},
    url = {https://www.etsi.org/deliver/etsi_gs/ZSM/001_099/003/01.01.01_60/gs_ZSM003v010101p.pdf}
}

@techreport{farrel2024framework,
  title={A framework for network slices in networks built from IETF technologies},
  author={Farrel, Adrian and Drake, John and Rokui, Reza and Homma, Shunsuke and Makhijani, Kiran and Contreras, Luis M and Tantsura, Jeff},
  year={2024},
  institution={RFC 9543, IETF, Tech. Rep. 9543}
}

@ARTICLE{arnhold.2024,
  author={Arnhold, Felipe and others},
  journal={IEEE Transactions on Network and Service Management}, 
  title={{Network Slicing Support by Fronthaul Interface in Disaggregated Radio Access Networks: A Survey}}, 
  year={2024},
  volume={21},
  number={4},
  pages={4510-4530},
  doi={10.1109/TNSM.2024.3400019}}

@article{rocha2026optimal,
  title={Optimal resource allocation with delay guarantees for network slicing in disaggregated RAN},
  author={Rocha, Fl{\'a}vio GC and Almeida, Gabriel M and Cardoso, Kleber V and Both, Cristiano B and De Rezende, Jos{\'e} F},
  journal={IEEE Transactions on Networking},
  year={2026},
  publisher={IEEE}
}

@article{ZANFERRARIMORAIS.2020,
    author = {Fernando Zanferrari Morais and others},
    title = {{When SDN meets C-RAN: A survey exploring multi-point coordination, interference, and performance}},
    journal = {Journal of Network and Computer Applications},
    volume = {162},
    pages = {102655},
    year = {2020}
}

@inproceedings{foukas2017orion,
  title={{Orion: RAN slicing for a flexible and cost-effective multi-service mobile network architecture}},
  author={Foukas, Xenofon and Marina, Mahesh K and Kontovasilis, Kimon},
  booktitle={Proceedings of the 23rd annual international conference on mobile computing and networking},
  pages={127--140},
  year={2017}
}

@book{dahlman20205g,
  title={5G NR: The next generation wireless access technology},
  author={Dahlman, Erik and Parkvall, Stefan and Skold, Johan},
  year={2020},
  publisher={Academic Press}
}

@article{svLarsen2018,
    author = {Larsen, Line and Checko, Aleksandra and Christiansen, Henrik},
    title = {{A Survey of the Functional Splits Proposed for 5G Mobile Crosshaul Networks}},
    year={2019},
    volume={21},
    number={1},
    pages={146-172},
    journal = {IEEE Communications Surveys \& Tutorials}
}

@article{checko2014cloud,
  title={{Cloud RAN for mobile networks—A technology overview}},
  author={Checko, Aleksandra and Christiansen, Henrik L and Yan, Ying and Scolari, Lara and Kardaras, Georgios and Berger, Michael S and Dittmann, Lars},
  journal={IEEE Communications surveys \& tutorials},
  volume={17},
  number={1},
  pages={405--426},
  year={2014},
  publisher={IEEE}
}

@article{saad2019vision,
  title={{A vision of 6G wireless systems: Applications, trends, technologies, and open research problems}},
  author={Saad, Walid and Bennis, Mehdi and Chen, Mingzhe},
  journal={IEEE network},
  volume={34},
  number={3},
  pages={134--142},
  year={2019},
  publisher={IEEE}
}

@article{lin2025bridge,
  title={{The bridge toward 6G: 5G-Advanced evolution in 3GPP Release I9}},
  author={Lin, Xingqin},
  journal={IEEE Communications Standards Magazine},
  volume={9},
  number={1},
  pages={28--35},
  year={2025},
  publisher={IEEE}
}

@ARTICLE{PlaceRAN.2023,
  author={Morais, Fernando Zanferrari and others},
  journal={IEEE Transactions on Mobile Computing}, 
  title={{PlaceRAN: Optimal Placement of Virtualized Network Functions in Beyond 5G Radio Access Networks}}, 
  year={2023},
  volume={22},
  number={9},
  pages={5434-5448}
}

@Article{Ornonez.2021,
    AUTHOR = {Ordonez-Lucena, Jose and others},
    TITLE = {{On the Rollout of Network Slicing in Carrier Networks: A Technology Radar}},
    JOURNAL = {Sensors},
    VOLUME = {21},
    YEAR = {2021},
    NUMBER = {23},
    ARTICLE-NUMBER = {8094}
}

@ARTICLE{_A2,
    author={Das, Sandip and Slyne, Frank and Ruffini, Marco},
    journal={IEEE Network}, 
    title={{Optimal Slicing of Virtualized Passive Optical Networks to Support Dense Deployment of Cloud-RAN and Multi-Access Edge Computing}}, 
    year={2022},
    volume={36},
    number={2},
    pages={131-138}
}

@article{_A4,
    author = {Sourav Mondal and Marco Ruffini},
    title = {{Optical Front/Mid-haul with Open Access-Edge Server Deployment Framework for Sliced {O-RAN}}},
    journal      = {CoRR},
    volume       = {abs/2110.09365},
    year         = {2021},
    url          = {https://arxiv.org/abs/2110.09365},
    eprint       = {2110.09365}
}

@ARTICLE{_A5,
    author={Ojaghi, Behnam and others},
    journal={IEEE Systems Journal}, 
    title={{SlicedRAN: Service-Aware Network Slicing Framework for 5G Radio Access Networks}}, 
    year={2022},
    volume={16},
    number={2},
    pages={2556-2567}
}

@INPROCEEDINGS{_A5.1,
    author={Ojaghi, Behnam and others},
    booktitle={IEEE International Conference on Communications (ICC)}, 
    title={{Sliced-RAN: Joint Slicing and Functional Split in Future 5G Radio Access Networks}}, 
    year={2019},
    pages={1-6}
}

@INPROCEEDINGS{_A7,
    author={Das, Sandip and Ruffini, Marco},
    booktitle={International Conference on Optical Network Design and Modeling (ONDM)}, 
    title={{Optimal virtual PON slicing to support ultra-low latency mesh traffic pattern in MEC-based Cloud-RAN}}, 
    year={2021},
    pages={1-5}
}

@inproceedings{_A8,
    author = {Budhdev, Nishant and others},
    title = {{FSA: Fronthaul Slicing Architecture for 5G Using Dataplane Programmable Switches}},
    year = {2021},
    booktitle = {Proceedings of the 27th Annual International Conference on Mobile Computing and Networking},
    pages = {723–735}
}

@INPROCEEDINGS{_A10,
    author={Das, Sandip and Ruffini, Marco},
    booktitle={Optical Fiber Communications Conference and Exhibition (OFC)}, 
    title={{PON Virtualisation with EAST-WEST Communications for Low-Latency Converged Multi-Access Edge Computing (MEC)}}, 
    year={2020},
    pages={1-3}
}

@ARTICLE{_A11,
    author={Das, Sandip and others},
    journal={Journal of Optical Communications and Networking}, 
    title={{Virtualized EAST–WEST PON architecture supporting low-latency communication for mobile functional split based on multiaccess edge computing}}, 
    year={2020},
    volume={12},
    number={10},
    pages={D109-D119}
}

@ARTICLE{_A13,
    author={Uzawa, Hiroyuki and others},
    journal={Journal of Optical Communications and Networking}, 
    title={{Dynamic bandwidth allocation scheme for network-slicing-based TDM-PON toward the beyond-5G era}}, 
    year={2020},
    volume={12},
    number={2},
    pages={A135-A143}
}

@ARTICLE{_A18,
    author={Li, Xi and others},
    journal={IEEE Communications Magazine}, 
    title={{5G-Crosshaul Network Slicing: Enabling Multi-Tenancy in Mobile Transport Networks}}, 
    year={2017},
    volume={55},
    number={8},
    pages={128-137}
}

@INPROCEEDINGS{_A19,
    author={Deiß, Thomas and others},
    booktitle={European Conference on Networks and Communications (EuCNC)}, 
    title={{Packet forwarding for heterogeneous technologies for integrated fronthaul/backhaul}}, 
    year={2016},
    pages={133-137}
}

@article{_A16,
    author = {González, Sergio and others},
    title = {{5G-Crosshaul: An SDN/NFV control and data plane architecture for the 5G integrated Fronthaul/Backhaul}},
    journal = {Transactions on Emerging Telecommunications Technologies},
    volume = {27},
    number = {9},
    pages = {1196-1205},
    year = {2016}
}

@inproceedings{ojaghi2019sliced,
  author={Ojaghi, Behnam and others},
  title={{Sliced-RAN: Joint slicing and functional split in future 5G radio access networks}},
  booktitle={IEEE International Conference on Communications (ICC)},
  pages={1--6},
  year={2019},
  organization={IEEE}
}

@article{ojaghi2022impact,
  author={Ojaghi, Behnam and others},
  title={{Impact of Network Densification on Joint Slicing and Functional Splitting in 5G}},
  journal={IEEE Communications Magazine},
  year={2022},
  publisher={IEEE}
}

\end{document}